\documentclass{article}

\usepackage{PRIMEarxiv}

\usepackage[utf8]{inputenc} 
\usepackage[T1]{fontenc}    
\usepackage{hyperref}       
\usepackage{url}            
\usepackage{booktabs}       
\usepackage{amsfonts}       
\usepackage{nicefrac}       
\usepackage{microtype}      
\usepackage{lipsum}
\usepackage{fancyhdr}       
\usepackage{graphicx}       
\graphicspath{{media/}}     
\usepackage{graphicx}  
\usepackage{epstopdf}
\usepackage{ragged2e}
\usepackage{algorithm}
\usepackage{algpseudocode}
\usepackage{enumerate}
\usepackage{subfigure}
\usepackage{amsmath}
\usepackage{bm}
\usepackage{stfloats}
\usepackage{siunitx}
\usepackage{booktabs}
\usepackage{multirow}
\usepackage[justification=centering]{caption}

\title{An Acoustic Signal Cavitation Detection Framework Based on XGBoost with Adaptive Selection Feature Engineering\thanks{\textit{Received 4 November 2021, Revised 7 February 2022, Accepted 10 February 2022, Available online 25 February 2022, Version of Record 26 February 2022.}}}

\author{
  Yu Sha \\
   Xidian University\\
   FIAS \thanks{\textit{FIAS: Frankfurt Institute for Advanced Studies}}\\
   XF-IJRC\thanks{\textit{XF-IJRC: Xidian-FIAS international Joint Research Center}}\\
   \And
  Johannes Faber \\
  FIAS \\
    \And
  Shuiping Gou \\
  Xidian University \\
    \And
  Bo Liu \\
  Xidian University \\
    \And
  Wei Li \\
  FIAS \\
    \And
  Stefan Schramm \\
  FIAS \\ 
    \And
  Horst Stoecker \\
  FIAS \\  
  Goethe Universit{\"a}t \\
  GSI\thanks{\textit{GSI: GSI Helmholtzzentrum f{\"u}r Schwerionenforschung GmbH}}\\
    \And
  Thomas Steckenreiter \\
  SAMSOM AG \\  
    \And
  Domagoj Vnucec \\
  SAMSOM AG \\  
    \And
  Nadine Wetzstein \\
  SAMSOM AG \\      
  \And
  Andreas Widl \\
  SAMSOM AG \\  
  \And
  Kai Zhou \thanks{\textit{Kai Zhou is the corresponding author. Email: zhou@fias.uni-frankfurt.de}}\\
  FIAS \\  
}

\begin{document}
\maketitle

\begin{abstract}
Valves are widely used in industrial and domestic pipeline systems. However, during their operation, they may suffer from the occurrence of the cavitation, which can cause loud noise, vibration and damage to the internal components of the valve. Therefore, monitoring the flow status inside valves is significantly beneficial to prevent the additional cost induced by cavitation. In this paper, a novel acoustic signal cavitation detection framework--based on XGBoost with adaptive selection feature engineering--is proposed. Firstly, a data augmentation method with non-overlapping sliding window (NOSW) is developed to solve small-sample problem involved in this study. Then, the each segmented piece of time-domain acoustic signal is transformed by fast Fourier transform (FFT) and its statistical features are extracted to be the input to the adaptive selection feature engineering (ASFE) procedure, where the adaptive feature aggregation and feature crosses are performed. Finally, with the selected features the XGBoost algorithm is trained for cavitation detection and tested on valve acoustic signal data provided by Samson AG (Frankfurt). Our method has achieved state-of-the-art results. The prediction performance on the binary classification (cavitation and no-cavitation) and the four-class classification (cavitation choked flow, constant cavitation, incipient cavitation and no-cavitation) are satisfactory and outperform the traditional XGBoost by $4.67\%$ and $11.11\%$ increase of the accuracy.
\end{abstract}


\section{Introduction}
Cavitation is a phenomenon of the dynamic process during which bubbles, also called cavities, form (later collapse) on the solid surface when the pressure at the contact point of a liquid and the solid surface is lower than the liquid's vapor pressure \cite{jazi2009detecting,zhao2007experimental}. When the bubble flows to the place where the liquid pressure exceeds the vapor pressure, the bubble collapses and the implosion instantaneously produces a great impact (shock waves) and high temperature \cite{brujan2002final,bonnier2002experimental}. Furthermore, severe pitting and wear can be induced on the solid surface \cite{song2014corrosion}.

Generally, cavitation can bring potential dangers to process plants, especially for valves, pumps, pipes or propellers. These potential dangers might quickly cause damage to components of the plants and loss of efficiency \cite{mckee2011review}. On the one hand, corrosion and destruction to valves, pumps or pipes can be caused by the cavitation occurrence \cite{schleihs20143d}, vibration of the internal structure and noise loudness levels might also increase \cite{yin2016numerical} due to the bubble collapse and rupture \cite{brennen2014cavitation}. On the other hand, the flow rate and fluid bulk properties inside the plants can be changed by the cavitation \cite{gholizadeh2012fluid,liu2005numerical} which further can lead to standstill of the process. Consequently, valves, pipes, pumps and related components are at the largest risk confronting cavitation. Many industries are fighting against cavitation, such as SAMSON AG for valves and other manufacturing plants \cite{clarke2011evaluating}. In the worst case, cavitation can lead to the closure of the factory, e.g. for test rack system with control valve. Therefore, it is vital to detect the cavitation of process plants (e.g. valves, pumps or pipes) at its early stages so as to ensure security and reduce any possible economic loss.

Cavitation is usually detected by comparing the fault and healthy conditions of specific devices under the monitored signals. According to the type of monitoring sensors, it can be divided into vibration signals based cavitation detection and acoustic signals based detection. Recently, some researchers have explored machine learning application for cavitation detection, based on features handcrafted from the vibration or acoustic signals.

Sakthivel et al. \cite{sakthivel2010vibration} extracted 11 statistical features from the vibration time domain signals. These features are then fed into the C4.5 decision tree \cite{quinlan2014c4} to classify the bearing fault, seal fault, impeller fault, bearing and impeller fault together with cavitation. Muralidharan et al.\cite{muralidharan2013feature} used the Continuous Wavelet Transform (CWT) \cite{rao2002wavelet} to replace the statistical feature extraction from vibration signals. Then the CWT achieved features are taken as input into the decision tree algorithm to for similar classification task as in \cite{sakthivel2010vibration}. In addition, Muralidharan et al. \cite{muralidharan2014fault} also studied the influence of different families and different levels of CWT for fault diagnosis of single-piece centrifugal oils using Support Vector Machine (SVM). Samanta et al. \cite{samanta2003artificial} extracted features from the original and pre-processed signals as the input of two different classifiers of the SVM and the artificial neural network (ANN) to identify normal and defective bearings. The parameters of the SVM and ANN are optimized by genetic algorithms \cite{holland1992genetic}, and the results explained the importance of feature selection to the classifier. Yang et al. \cite{yang2005cavitation} extracted 4 statistical features from the vibration time domain signals as the input of SVM to detect cavitation and no cavitation of the butterfly valve. Bordoloi et al. \cite{bordoloi2017identification} proposed a SVM method using directly the vibration signal data of bearing block and pump casing to diagnose blockage level and cavitation intensity. Panda et al. \cite{panda2018prediction} extracted statistical features from the time domain vibration signal of the pump as the input of SVM to distinguish cavitation and flow block. Rapur et al. \cite{rapur2018automation} proposed an intelligent detection method based on SVM to classify mechanical fault and flow rate using combination of motor line current and pump vibration signal as input. Shervani-abar. \cite{shervani2018cavitation} proposed a multi-class cavitation detection method based on the vibration signal of the axial flow pump using SVM.

Zouari et al. \cite{zouari2004fault} proposed a vibration signal fault detection method for centrifugal pumps using neural network and neuro-fuzzy technology. Rajakarunakaran et al. \cite{rajakarunakaran2008artificial} proposed a centrifugal pump fault detection using a feedforward network and a binary adaptive resonance network (ART1). Siano et al. \cite{siano2018diagnostic} proposed a method combining ANN and nonlinear regression to diagnose cavitation of time domain vibration signals. Nasiri et al. \cite{nasiri2011vibration} extracted features from the vibration signal of the centrifugal pump as the input of the neural network to detect cavitation. Jia et al. \cite{jia2016deep} proposed a deep neural network (DNN) to directly extract features from the original rolling element bearings and planetary gearboxes data set for fault diagnosis. Zhao et al. \cite{zhao2016fault} proposed a deep learning method to extract features from non-stationary vibration signals and diagnose centrifugal pump faults. Tiwari et al. \cite{tiwari2021blockage} extracted 6 statistical features from the time domain pressure data, and then these features are input to the neural work to detect blockage and cavitation. Potocnik et al. \cite{potovcnik2021condition} extracted spectral and psychoacoustic features from the valve acoustic data and then these features are input to a variety of machine learning algorithms to classify the cavitation, flow noise, whistling and rattling.

It can be found from the literature review that, many traditional machine learning methods (like, decision tree, SVM or ANN) have been explored for cavitation or fault detection. However, they mostly take vibration signal and also they may not be able to capture sensitive features of the signal. On the other hand, although deep learning can do end-to-end cavitation detection to solve the problem of capturing underlying sensitive features, it is usually limited by lack of interpretability and stability in practice. In addition, cavitation detection based on acoustic signals is a more challenging task compared to the detection with vibration signals due to the sensor quality and position requirment. Accordingly, in this paper we propose an acoustic signal cavitation detection framework using XGBoost together with adaptive selection feature engineering. The main contributions of this paper are as follows:
\begin{itemize}
    \item The XGBoost, as an ensemble learning algorithm with good stability, is applied to the cavitation detection on acoustic signals for the first time.
    \item In order to tackle the small-sample problem, the non - overlapping sliding window data augmentation method is introduced in the study.
    \item Three types of statistical features (central trend, dispersion degree and distribution shape) are extracted for the task, which shows good sensitivity for cavitation identification.
    \item The adaptive selection feature engineering, including feature aggregation and feature crosses, is proposed to increase the expression ability of the features, and we demonstrated that this can further increase the sensitivity of the features for the task.
\end{itemize}

The remainder of this paper is organized as follows. Section \ref{sec:2-Data Acquisition} introduces the data acquisition process for control valve cavitation detection. Section \ref{sec:3-methods} firstly describes the data augmentation method based on non - overlapping sliding window. Then, statistical feature extraction and the adaptive selection feature engineering are introduced. Finally, the XGBoost based cavitation detection framework on acoustic signals with adaptive selection feature engineering is introduced. Section \ref{sec:4-Results and Discussion} presents the classification performance of our method. Conclusion are presented in Section \ref{sec:5-Conclusion}.

\section{Experimental setup and data acquisition}
\label{sec:2-Data Acquisition}
\begin{figure*}
    \centering
    \includegraphics[width=0.9\textwidth,height=50mm]{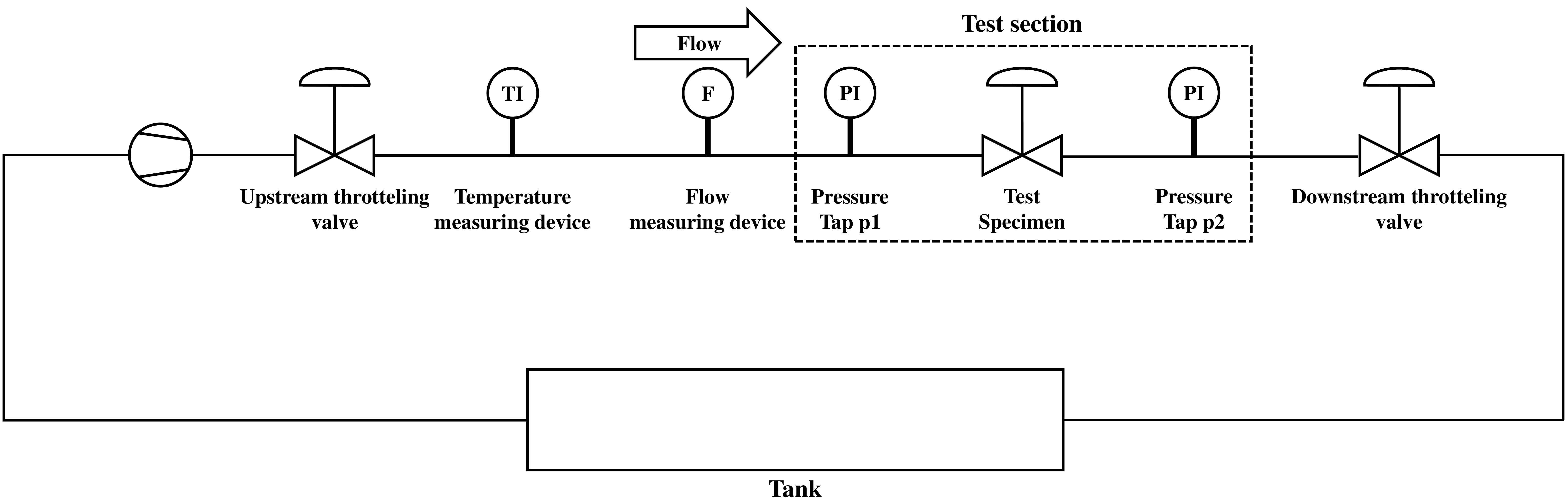}
    \caption{\centering{view of the test rack at SAMSON AG (Figure provided by SAMSON AG) for the data acquisition process.}}
    \label{fig:system}
\end{figure*}

SAMSON AG devised a test rack of control valve (SAMSON AG type 3241, DN 80, PN40, Kvs 25 with positioner type 3730-6) with running water as the flowing medium inside to generate different flow status by gauging operation conditions accordingly: upstream pressure, downstream pressure and the valve stroke (see Figure \ref{fig:system}). The test bench is equipped with a set of sensors to measure the temperature of the test medium $T$, the upstream pressure ${p_1}$, downstream pressure ${p_2}$ and vapor pressure of fluid ${p_v}$ and the volumetric flow rate $Q$. Additionally, the test valve mounted inside the bench measures the absolute valve stroke $h$ and the sound intensity ${L_p}$ with a special sensor directly mounted on the valve body. Furthermore, the test bench includes two additional control valves upstream and downstream of the test valve. A control system controls the pumps as well as these two valves to modify the total volumetric flow through the test bench. The two additional valves are used to influence the upstream and downstream pressure around the test valve. All sensors and the test valve are mounted between these two external control valves (more details see Table \ref{tab:sensors}). The pipes between the external valves and the test valve are long enough to ensure an undisturbed flow for proper measurements. Water, which can also be heated, was used as the test medium for all tests. The water temperature was maintained at 25 – 40 $^{\circ}$C in order to hold the vapor pressure nearly constant and to eliminate the influence of temperature on the cavitation.

Two piezo elements were introduced to record directly the structure-borne noise of the valve: one placed on the body of the control valve and the other one to the NAMUR at the bonnet. Furthermore two microphones were placed in a distance of 1m to the control valve to record the airborne noise in a frequency range between 40 Hz up to 20 kHz: one high-end microphone as well as a low cost microphone which typically is used in smartphones. Figure \ref{fig:system} shows a schematic view of the experimental setup, where the data is generated. 

The test section in figure \ref{fig:system} of is in a separate room and all other noise sources from the plant like the pump and the upstream and downstream throttling valves are in another room (below the best section room, in the cellar). Therefore, there is a spatial separation of the test section to all possible noise sources. Also bellow expansion joints are integrated in the pipe system in order to the transport of vibrations into the test section. All these measures ensure that a surrounding noise of maximum 55 dB were present during the tests. The preparation of the different states is very simple. The principle procedure is described in the standard IEC 60534-8-2. The considered states depend on the differential pressure ratio ($({p}_{1}-{p}_{2})/({p}_{1}-{p}_{v})$) of the valve. Therefore, the valve opening will be held constant during the test. The temperature of the medium is during the test constant, which means the vapor pressure will be held constant. Furthermore, the upstream pressure of the valve will also be held constant (e.g. 10 bar (a)) with the process control system of the plant. Therefore, in the consequence the different states will be adjusted by varying the downstream pressure in steps of 0.1,…,0.4 bar starting at 1bar (a). For each step the downstream pressure will be adjusted with the process control system of the plant and based on the steady state of the flow the noise measurement concerning the adjusted step will be started. As a result, there is for a constant valve opening and a constant valve upstream pressure a noise emission characteristic (noise emission versus differential pressure ratio). And based on the noise emission characteristic the different states will be determined. This extensive procedure for determination of cavitation states cannot be realized in customers plants.

\newcommand{\tabincell}[2]{\begin{tabular}{@{}#1@{}}#2\end{tabular}}  
\begin{table*}[htbp]
\caption{The detailed information of various measurement sensors}
\label{tab:sensors}
\centering
\begin{tabular}{llcc}
\toprule
Physical parameters                                                                   & Sensor types       & Measuring ranges & Tolerances \\ 
 \midrule
Upstream pressure                                                                     & SAMSON Type 6054   & 0-16 bar          & ± 0.09 bar \\
Downstream pressure                                                                   & SAMSON Type 6054   & 0-16 bar         & ± 0.09 bar \\
Flow rate                                                                             & Krohne Type M950   & 0-180 m³/h       & ± 0.9 m³/h \\
Valve stroke                                                                          & Sylvac Type s229   & 0-50 mm          & ± 0.001 mm \\
Medium temperature                                                                    & SAMSON Type 5204   & -20-150 $^{\circ}$C       & ± 0.45 $^{\circ}$C  \\
\begin{tabular}[c]{@{}l@{}}Structure-borne noise \\ (at valve body)\end{tabular}      & Vallen Type VS45-H & 20-450 kHz       & \begin{tabular}[c]{@{}c@{}}Sensitivity -63 dB  re 1V/µbar\\ (Accuracy according to individual calibration)\end{tabular} \\
\begin{tabular}[c]{@{}l@{}}Structure-borne noise \\ (at valve body)\end{tabular}      & PCB Type M353 B17  & 0-30000 Hz       & Sensitivity 10 mV/g, Accuracy ± 3 dB    \\
\begin{tabular}[c]{@{}l@{}}Structure-borne noise \\ (at NAMUR at bonnet)\end{tabular} & PCB Type M353 B17  & 0-30000 Hz       & Sensitivity 10 mV/g, Accuracy ± 3 dB    \\ 
\bottomrule
\end{tabular}
\end{table*}

By varying the differential pressure at various constant upstream pressures of the control valve different operation conditions were adjusted, five flow status are induced in the acoustic signal data: cavitation choked flow, constant cavitation, incipient cavitation, turbulent flow and background no-flow. The status turbulent flow is a flow through the control valve without any cavitation noise. Starting at a certain differential pressure ratio within a certain range only a few vapor bubbles will be generated and the implosion of these bubbles is causing an increase of the noise emission, which is defined as incipient cavitation. By increasing the concentration of vapor bubbles also the noise emission is increasing up to a noise maximum which is defined as constant cavitation. Achieving the noise maximum also the concentration of the vapor bubbles is maximum which leads to a choked flow condition with cavitation and a decreasing noise behavior. To obtain the data in this study, the noise is measured based on the IEC 60534-8-2.

The recorded acoustic signals are time series with sampling rate 1562.5 kHz and time duration of 3 s, so the number of entries of each signal sample is 4687500. In total, 356 samples are analyzed in this paper with partition into different flow status listed in Table \ref{tab:data}. For each flow state, we control seven different valve opening rates (100$\%$, 90$\%$, 75$\%$, 25$\%$, 10$\%$ and 5$\%$) to obtain acoustic signals at each upstream pressure (10, 9, 6 and 4), respectively. And the parameters of the valve opening rate and the upstream pressure are set by SAMSON AG's professional engineers based on practical experience.

\begin{table*}[htbp]
    \centering
    \caption{Different upstream pressure and valve opening in five flow status}
    \begin{tabular}{ccccc}
	\toprule 
	\multicolumn{2}{c}{Flow status}&  \tabincell{c}{Upstream pressure \\($bar(a)$)}& \tabincell{c}{Valve opening\\($\%$)} &\tabincell{c}{ Number of \\samples} \\ 
	\midrule
	\multirow{3}{*}{Cavitation} 
	 & Cavitation choked flow & [10,9,6,4] &[100,90,75,50,25,10,5] & 72\\
	 & Constant cavitation & [10,9,6,4] &[100,90,75,50,25,10,5] & 93 \\
	 & Incipient cavitation & [10,9,6,4] &[100,90,75,50,25,10,5] & 40 \\
	 \midrule
	 \multirow{2}{*}{No cavitation} 
	 & Turbulent flow & [10,9,6,4] &[100,90,75,50,25,10,5] & 118 \\
	 & No flow & [10,9,6,4] &[100,90,75,50,25,10,5] & 33 \\
	\bottomrule
    \end{tabular}
    \label{tab:data}
\end{table*}

\section{Methods}
\label{sec:3-methods}
This section presents our XGBoost based cavitation detection framework on acoustic signals. Firstly, a data augmentation method with the aid of non-overlapping sliding window is proposed for tackling the small-sample issue. Secondly, the statistical features are calculated from the frequency domain signal. Thirdly, adaptive selection feature engineering (ASFE) is proposed. Finally, the cavitation detection framework using XGBoost on acoustic signals with adaptive selection feature engineering is explained.

\subsection{Data Augmentation}
In general, machine learning is driven by big data \cite{l2017machine}. However, our data set has only 356 measured acoustic signals, so data augmentation is essential for handling this kind of small-sample problem. Data augmentation can improve the accuracy of the model and prevent over-fitting, and it can also teach the model the desired invariance and robustness properties \cite{shorten2019survey, Pang:2016vdc}. 

Considering the purposed maintaining for steady flow status (i.e. for each measurement it is the same fluid status class within the 3-second record duration) in each recorded data sample and the fine resolution for the sensor, one can actually split every sample into several pieces with each still holding enough essential information to decipher the flow status, but also with independent characteristics per piece due to the intrinsic randomness of the noise emission - given the piece is not so short. Therefore, we propose here a data augmentation method based on Non - Overlapping Sliding Window (NOSW), see Figure \ref{fig:sliding-window}. The method is divided into two steps:
\begin{figure}[htbp]
    \centering
     \includegraphics[width=0.5\textwidth,height=55mm]{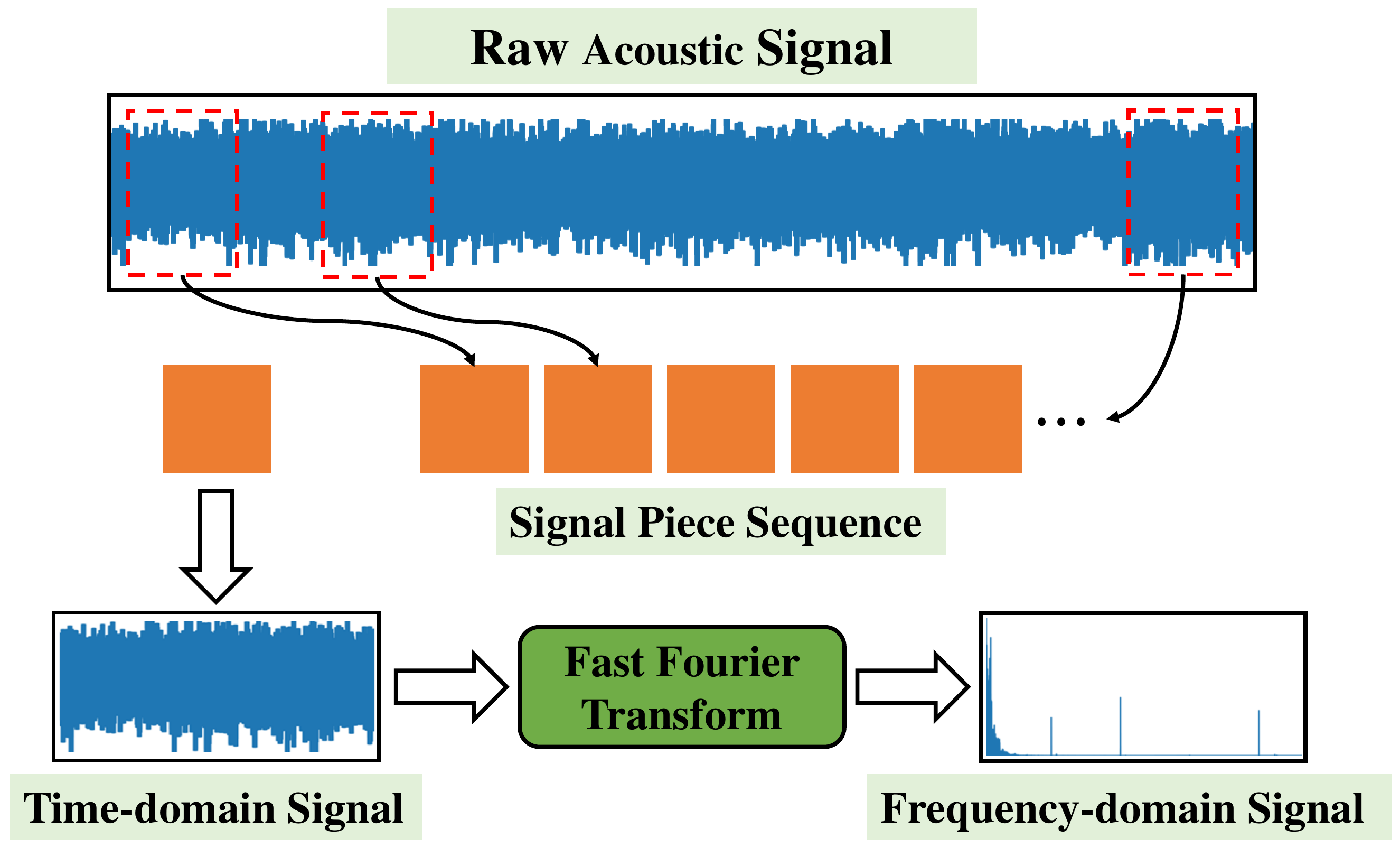}
    \caption{Data augmentation based on non-overlapping sliding window.}
    \label{fig:sliding-window}
\end{figure}
\begin{enumerate}[Step 1:]
\item The raw acoustic signal is split into pieces by a non-overlapping sliding window.
\item The time-domain signal data is then transformed into frequency-domain data by Fast Fourier Transform (FFT).
\end{enumerate}

Table \ref{tab:window-size} shows the number of samples in training and testing sets under different sized sliding windows.


\begin{table}[htbp]
\centering
\caption{The number of training and testing set under different window sizes}
\begin{tabular}{ccc}
\toprule
      \multirow{2}{*}{\tabincell{c}{Window \\Size}} & \multicolumn{2}{c}{Number} \\ 
      \cline{2-3}
      & Training Set & Testing set   \\
      \midrule
       2334720 & 568    & 144  \\
       1556480 & 852    & 216  \\
       1167360 & 1136   & 288  \\
       933888  & 1420   & 360  \\
       778240  & 1704   & 432  \\
       667062  & 1988   & 504  \\
       583680  & 2272   & 576  \\
       518825  & 2556   & 648  \\
       466944  & 2840   & 720  \\
\bottomrule
\end{tabular}
\label{tab:window-size}
\end{table}

\subsection{Feature Extraction}
\label{sec:featureExtracion}
Although the frequency-domain signal contains abundant information, it also contains non-essential information e.g. the background noise. A statistical feature extraction thus can further bring in effectiveness for cavitation detection under statistical learning paradigm. In general, the statistical features are the average attribute of the data those can be captured. Many of these statistical features are based on moments and therefore the method of estimating these statistical features are reflecting their relationship with the distribution of the random variable. Normally, the probability density function can be decomposed into the components of moments. 

In this paper, three types of statistical features are extracted: central trend statistics, dispersion degree statistics and distribution shape statistics, given a frequency-domain signal ${x_i}(i = 1, \ldots ,n)$ with $n$ the number of data points.

\textit{\textbf{Central Trend Statistics}} The central trend statistics represent the central tendency of the overall distribution and is thus a crucial feature.  In this paper, the central trend statistics include mean value $\mu $, median $M$, low quartile ${Q_1}$ and upper quartile ${Q_3}$, which are defined as follows respectively:
\begin{equation} 
\label{eq:CenTrendStatistics}
 \begin{split}
  & \mu  = \frac{1}{n}\sum\limits_{i = 1}^n {{x_i}},\\
  & M= \left\{ \begin{array}{l}{x_{{{(n + 1)} \mathord{\left/{\vphantom {{(n + 1)}2}}\right.\kern-\nulldelimiterspace}2}}},\,n \in 2n + 1,\\\frac{{{x_{({n\mathord{\left/{\vphantom {n 2}} \right.\kern-\nulldelimiterspace} 2})}} + {x_{(({n \mathord{\left/{\vphantom {n 2}} \right.\kern-\nulldelimiterspace} 2}) +1)}}}}{2},\,n \in 2n,\end{array}\right.\\
  & {Q_p} = \left\{ \begin{array}{l}{x_{(\left[ {np} \right] + 1)}},\,np \notin \mathbb{Z},\\\frac{1}{2}[{x_{(np)}}{x_{(np + q)}}],\,np \in \mathbb{Z}.\end{array} \right.\\
 \end{split}
\end{equation}
where, ${x_i},i = 1, \ldots ,n$ is a frequency domain signal data, $n$ is the number of the data points, $p = 1,3$ and $\mathbb{Z}$ represents an integer.

\textit{\textbf{Dispersion Degree Statistics}} The dispersion degree statistics is the distribution of all observations off-center. In this paper, the dispersion degree statistics include minimum $Min$, maximum $Max$, inter quartile range $IQR$, standard deviation $\sigma$, root mean square $RMS$ and square root amplitude $SRA$, which are defined as follows respectively:
\begin{equation} 
\label{eq:DispersionDegreeStatistics}
 \begin{split}
  & IQR = {Q_3} - {Q_1},\\
  & \sigma  = \sqrt {\frac{1}{n}\sum\limits_{i = 1}^n {{{({x_i} - \mu )}^2}} },\\
  & RMS = \sqrt {\frac{{\sum\nolimits_{i = 1}^n {x_i^2} }}{n}},\\
  & SRA = {(\frac{1}{n}\sum\limits_{i = 1}^n {\sqrt {\left| {{x_i}} \right|} } )^2}.\\
 \end{split}
\end{equation}
where, $\mu$ is mean, ${Q_1}$ is low quartile, ${Q_3}$ is upper quartile, ${x_i},i = 1, \ldots ,n$ is a frequency-domain signal data and $n$ is the number of the data points.

All data is in $[Min,Max]$, the number of data in $[Min,{Q_1}]$, $[{Q_1},M],\,[M,{Q_3}],\,[{Q_3},Max]$ accounts for about a quarter. When the interval is short, it means that the points in the interval are concentrated, and vice versa.

If the $M$ is located in the middle of ${Q_1}$ and ${Q_2}$, the data distribution is more symmetrical. If the distance between $Min$ and $M$ is greater than the distance between $Max$ and $M$, the data distribution is inclined to the left, otherwise the data distribution is inclined to the right. And we can see the length of the tail of the distribution. 

\textit{\textbf{Distribution Shape Statistics}} The distribution shape statistics is a crucial feature to measure the distribution shape of observation data. In this paper, the distribution shape statistics include kurtosis $\kappa$, skewness $\chi$, shape factor ${s_f}$, clearance factor ${c_{lf}}$ and crest factor ${c_f}$, which are defined as follows:
\begin{equation} 
\label{eq:DistributionShapeStatistics}
 \begin{split}
  &\kappa  = \frac{{\frac{1}{n}\sum\nolimits_{i = 1}^n {{{({x_i} - \mu )}^4}} }}{{{\sigma ^4}}},\\
  & \chi  = \frac{{\frac{1}{n}\sum\nolimits_{i = 1}^n {{{({x_i} - \mu )}^3}} }}{{{\sigma ^3}}},\\
  & {s_f} = \frac{{RMS}}{{\frac{1}{n}\sum\limits_{i = 1}^n {\left| {{x_i}} \right|} }},\\
  & {c_{lf}} = \frac{{\max \{ \left| {{x_i}} \right|\} }}{{SRA}},\\
  & {c_f} = \frac{{\max \{ \left| {{x_i}} \right|\} }}{{RMS}}.\\
 \end{split}
\end{equation}
where, $\mu$ is mean, $\sigma$ is standard deviation, $SRA$ is square root amplitude, $RMS$ is root mean square, ${x_i},i = 1, \ldots ,n$ is a frequency-domain signal data, $n$ is the number of the data points.

The kurtosis $\kappa$ \cite{mardia1970measures} is a statistic that describes the steepness of the distribution of all observations in the overall sample. And it is defined as the standardized 4rd central moment of the sample. If $\kappa  = 0$, it means that the overall data distribution is the same as the normal distribution. If $\kappa  > 0$, it means that the overall data distribution is steeper compared with the normal distribution. If $\kappa  < 0$, it means that the overall data distribution is flatter compared with the normal distribution. 

The skewness $\chi$ \cite{mardia1970measures} is similar to the kurtosis $\kappa$. The skewness $\chi$ is a statistic that describes the symmetry of the distribution of all observations in the overall sample. And it is defined as the standardized 3rd central moment. If $\chi = 0$, it means that the overall data distribution is the same as the normal distribution. If $\chi \ge 0$, it means that the overall data distribution is positively skewed compared with the normal distribution. If $\chi \le 0$, it means that the overall data distribution is negatively skewed compared with the normal distribution.

Shape factor ${s_f}$ is the ratio of $RMS$ to average rectified value. Clearance factor ${c_{lf}}$ is the ratio of signal peak to $SRA$. Crest factor ${c_f}$ is the ratio of signal peak to $RMS$. These features have similar physical meanings and are often used for fault detection.

\subsection{Adaptive Selection Feature Engineering (ASFE)}
Feature engineering is an important part of data mining. In general, the upper limit of machine learning is determined by data and features. And the upper limit is approached by algorithms and models \cite{turner1999conceptual,nargesian2017learning}. In addition, deep learning has achieved good results in many fields. The main reason is that it has a good ability of feature extraction and feature processing. How to increase the nonlinear relationship between features is very necessary. Therefore, the ASFE is proposed to enhance the expression ability of original features, i.e., to enhance the quality of the features. The method includes two parts: Adaptive Feature Aggregation and Adaptive Feature Crosses.

\subsubsection{Adaptive Feature Aggregation} 
\label{sec:AdaptiveFeatureAggregation} 
\begin{figure*}
    \centering
    \includegraphics[width=0.85\textwidth,height=60mm]{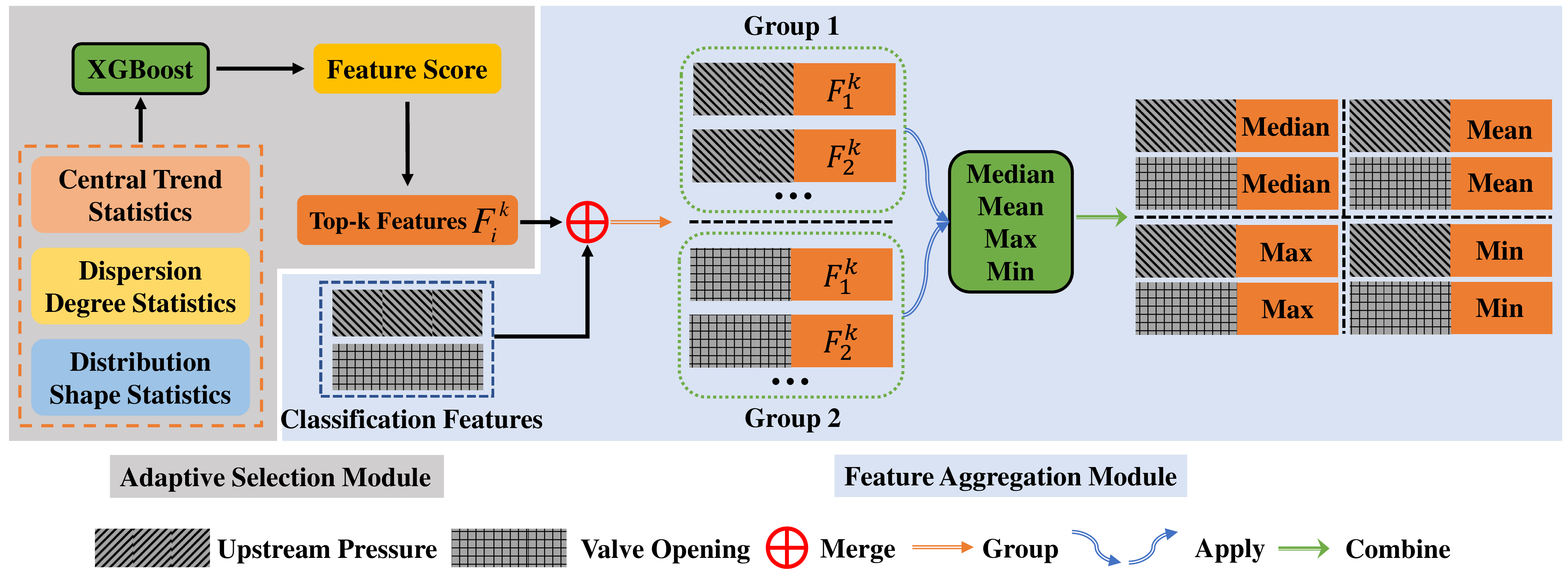}
    \caption{The diagram of the adaptive feature aggregation. The adaptive feature aggregation include the adaptive selection module and the feature aggregation module.}
    \label{fig:featureaggregation}
\end{figure*}

The adaptive feature aggregation is to aggregate numerical features according to classification features. The adaptive feature aggregation includes two modules: Adaptive Selection Module and Feature Aggregation Module, see Figure \ref{fig:featureaggregation}. 

\textbf{Adaptive Selection Module} Firstly, the numerical features are input to XGBoost. Where, the numerical features include the central trend statistics, the dispersion degree statistics and the distribution shape statistics. Then, the score of each feature is obtained. Finally, the $top - k(k = 5, \cdots ,10)$ features $F_i^k(i = 1, \cdots ,k)$ are selected for feature aggregation according to the feature scores. 

\textbf{Feature Aggregation Module} The module includes three operations: \textit{Group}, \textit{Apply} and \textit{Combine}.

\textbf{\textit{-Group}} The upstream pressure and valve opening are classification features. Therefore, the numerical features are grouped according to the upstream pressure and valve opening. Then, the group of upstream pressure $Group-1$ and the group of valve opening $Group-2$ are obtained. Where, the statistical features of $Group-1$ and $Group-2$ are respectively denoted as $F_i^k(p)$, $F_i^k(o)$. And $F_i^k(p) \cup F_i^k(o) = F_i^k$.

\textbf{\textit{-Apply}} The four operations are respectively applied to $Group-1$ and $Group-2$: Median, Mean, Maximum, and Minimum.

\textbf{\textit{-Combine}} The $Group-1$ and the $Group-2$ are combined according to the same type of application.

The aggregation features $F(A)$ are generated through the adaptive feature aggregation. The number of aggregation features $F(A)$ are $Num(F(A)) = Num(apply) \times k$. Table \ref{tab:featureaggregation} shows the number of aggregation features under different $k$ through adaptive feature aggregation.
\begin{table}[htbp]
    \centering
    \caption{The number of aggregation features under different top $k$ by adaptive feature aggregation.}
    \begin{tabular}{cccc}
	\toprule 
	& $k$  & The number of aggregation features   \\
	\midrule
	&5 & $4 \times 5 = 20$\\
	&6 & $4 \times 6 = 24$\\
	&7 & $4 \times 7 = 28$\\
	&8 & $4 \times 8 = 32$\\
	&9 & $4 \times 9 = 36$\\
	&10 & $4 \times 10 = 40$\\
	\bottomrule
    \end{tabular}
    \label{tab:featureaggregation}
\end{table}

\subsubsection{Adaptive Feature Crosses}
\label{sec:AdaptiveFeatureCrosses} 
\begin{figure}[htbp]
    \centering
    \includegraphics[width=0.50\textwidth,height=45mm]{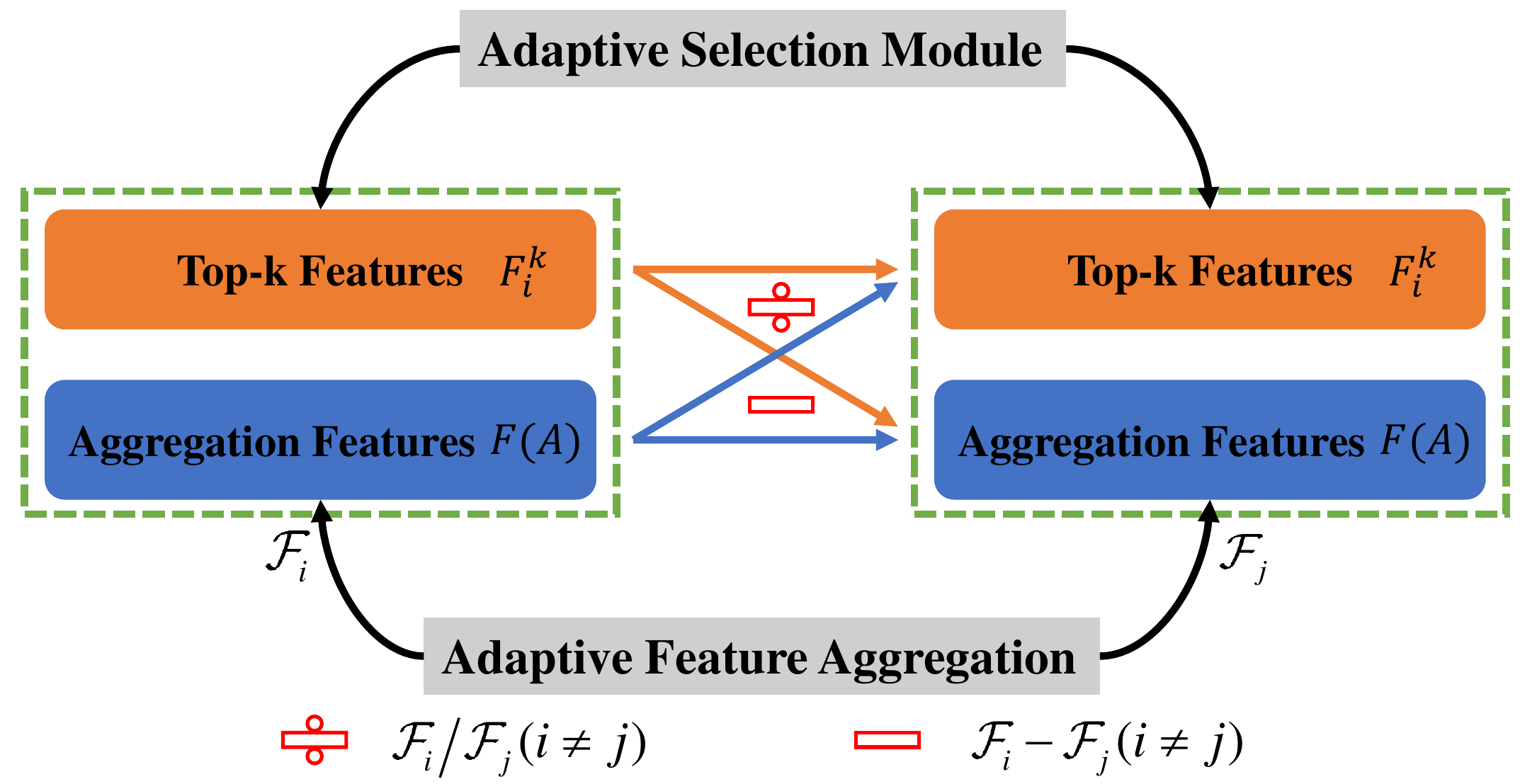}
    \caption{The diagram of the adaptive feature crosses.}
    \label{fig:featurecross}
\end{figure}

The adaptive feature crosses is a cross feature formed by crossing two or more features. Similar to the adaptive feature aggregation, the adaptive feature crosses is also composed of two modules: Adaptive Selection Module (see \ref{sec:AdaptiveFeatureAggregation}) and Feature Crosses Module, see Figure \ref{fig:featurecross}. 

Since the aggregation features $F(A)$ are used in the adaptive feature crosses, the adaptive feature crosses is after the adaptive feature aggregation. Firstly, the $top-k$ features $F_i^k$ and the aggregation features $F(A)$ are obtained through the Adaptive Selection Module and the Adaptive Feature Aggregation, respectively. Then, the $top-k$ features $F_i^k$ and the aggregation features $F(A)$ are merged and denoted as $\mathcal{F}_i(i = 1, \cdots ,Num(F(A) )+ k)$. Thirdly, $\mathcal{F}_i$ are copied and denoted as $\mathcal{F}_j(j = 1, \cdots ,Num(F(A) )+ k)$. Finally, ${{{\mathcal{F}_i}} \mathord{\left/{\vphantom {{{\mathcal{F}_i}} {{\mathcal{F}_j}}}} \right.\kern-\nulldelimiterspace} {{\mathcal{F}_j}}}$ and ${\mathcal{F}_i} - {\mathcal{F}_j}$, where $i \ne j$.

The number of cross features $F(C)$ are $Num(F(C)) = 2 \times (k + Num(F(A))) \times (k + Num(F(A)) - 1)$. Table \ref{tab:featurecross} shows the number of cross features under different $k$ through adaptive feature crosses.
\begin{table}[htbp]
    \centering
    \caption{The number of cross features under different top $k$ by adaptive feature cross.}
    \begin{tabular}{cccc}
	\toprule 
	& $k$  & The number of cross features   \\
	\midrule
	&5 & 1200\\
	&6 & 1300\\
	&7 & 1404\\
	&8 & 1512\\
	&9 & 1624\\
	&10 & 1740\\
	\bottomrule
    \end{tabular}
    \label{tab:featurecross}
\end{table}

\subsection{eXtreme Gradient Boosting with adaptive selection feature engineering}
eXtreme Gradient Boosting (XGBoost) \cite{chen2015xgboost} is a supervised ensemble learning algorithm. It uses the second order Taylor \cite{levenberg1944method} expansion loss function with further a regularization term added based on the Gradient Boosting Decision Tree (GBDT) \cite{friedman2001greedy} algorithm. It has been widely applied in different kinds of data mining fields. 

For a given data set with $n$ samples and $m$ features, ${\cal D} = \{ ({x_i},{y_i})\} (\left| {\cal D} \right| = n,{x_i} \in {\mathbb{R}^m},{y_i} \in\mathbb{R})$, a tree ensemble model uses $K$ additive functions is used to predict the output,
\begin{equation}
{{\hat y}_i} = \sum\limits_{k = 1}^K {{f_k}({x_i}),{f_k} \in {\cal F},} 
\end{equation}
where ${\cal F} = \{ f(x) = {w_{q(x)}}\} (q:{\mathbb{R}^m} \to T,w \in {\mathbb{R}^T})$ is the space of Classification and Regression Trees (CART)\cite{loh2011classification}. $q$ is an independent tree structure. $T$ is the number of leaves in each tree. $w$ is the weight of the leaf node. Therefore, the regularized objective function of XGBoost is defined as follows:
\begin{equation}
\label{eq:objectivefunction}
O{b_j} = \sum\limits_i {l({{\hat y}_i},{y_i}) + \sum\limits_k {\Omega ({f_k})} } ,
\end{equation}
\begin{equation}
\label{eq:regularizeditem}
\Omega ({f_k}) = \gamma T + \frac{1}{2}\lambda \sum\limits_{j = 1}^T {w_j^2} ,
\end{equation}
where $\sum\limits_i {l({{\hat y}_i},{y_i})}$ is the normal loss function with ${{y_i}}$ the actual value and ${{{\hat y}_i}}$ the predicted value. ${\sum\limits_k {\Omega ({f_k})} }$ is the regularization term that is composed of the sum of the complexity of all classification and regression trees. $T$ is the number of leaves in each tree. $\gamma$ is the penalty coefficient, which is used to control the number of leaf nodes. ${w_j}$ is the weight of the leaf node. The regularization term (\ref{eq:regularizeditem}) helps to smooth the final learned weights and to avoid over-fitting.

When a new CART are generated, XGBoost needs to fit the error of the previous prediction. When $t$ CART are generated, that is, after $t$-th iterations, the objective function becomes
\begin{equation}
Ob_j^{(t)} = \sum\limits_{i = 1}^n {l[{y_i},\hat y_i^{(t - 1)} + {f_t}({x_i})] + } \Omega ({f_t}),
\end{equation}
where $l[{y_i},\hat y_i^{(t - 1)} + {f_t}({x_i})]$ represents the error function and $\Omega ({f_t})$ represents the regularization term.

In order to optimize the objective function quickly, we approximate the objective function using the second-order Taylor expansion,
\begin{equation}
Ob_j^{(t)} \cong \sum\limits_{i = 1}^n {l[{y_i},{{\hat y}^{(t - 1)}} + {g_i}{f_t}({x_i}) + \frac{1}{2}{h_i}f_t^2({x_i})] + } \Omega ({f_t}),
\end{equation}
where ${g_i} = {\partial _{{{\hat y}^{(t - 1)}}}}l({y_i},{{\hat y}^{(t - 1)}})$ is the first order gradient statistics on the objective function, ${h_i} = \partial _{{{\hat y}^{(t - 1)}}}^2l({y_i},{{\hat y}^{(t - 1)}})$ is the second order gradient statistics on the objective function. ${{\hat y}^{(t - 1)}}$ represents the prediction result of the previous $t-1$ tree, and ${{f_t}({x_i})}$ represents the model of the $i$-th tree.

Since the objective function only depends on the first and second order stochastic gradients of the loss function over each data point, the constant term in the objective function can be removed, thus the following simplified objective function is used at each training step $t$,
\begin{equation}
\label{eq:SimplifiedObjFunc}
Ob_j^{(t)} \cong \sum\limits_{i = 1}^n {[{g_i}{f_t}({x_i}) + \frac{1}{2}{h_i}f_t^2({x_i})] + \Omega } ({f_t}).
\end{equation}

We substitute Equation (\ref{eq:regularizeditem}) into Equation (\ref{eq:SimplifiedObjFunc}), and arrived at the final objective function as follows :
\begin{equation}
Ob_j^{(t)} = \sum\limits_{j = 1}^T {[{G_j}{w_j} + \frac{1}{2}({H_j} + \lambda )w_j^2]}  + \gamma T,
\end{equation}
where ${G_j} = \sum\limits_{i \in {I_j}} {{g_i}}$ and ${H_j} = \sum\limits_{i \in {I_j}} {{h_i}}$. ${I_j} = \{ i|q({x_i}) = j\}$ is defined as the instance set of leaf $j$.

\begin{figure}[htbp]
    \centering
    \includegraphics[width=0.45\textwidth,height=205mm]{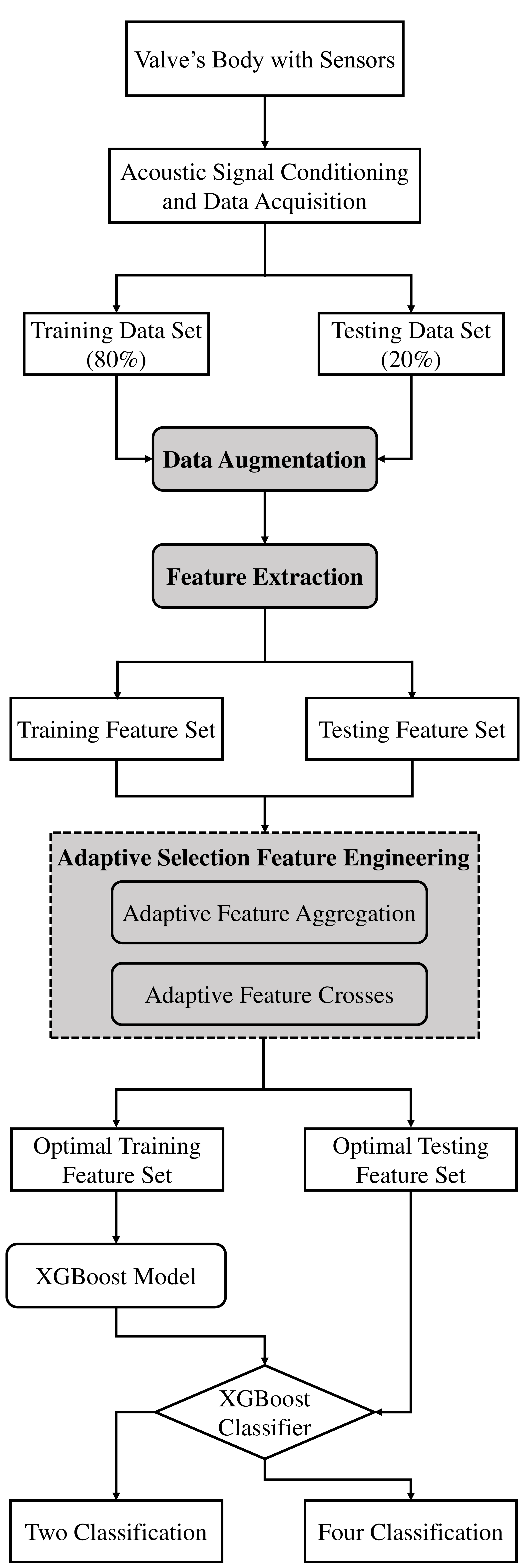}
    \caption{Flow chart of our cavitation detection framework on acoustic signal, based on XGBoost with the adaptive feature engineering.}
    \label{fig:framework}
\end{figure}

Based on the above XGBoost together with the adaptive selection feature engineering, we developed the cavitation detection framework on the valve acoustic signal, see Figure \ref{fig:framework} for the flow chart. Our method mainly includes five modules: data acquisition, data augmentation, feature extraction, adaptive selection feature engineering and XGBoost classification. And our source code is released at
\url{https://github.com/CavitationDetection/XGBoost_ASFE}.

First, in the data acquisition module we collected the acoustic data of the valve (see \ref{sec:2-Data Acquisition}) at different flow status. A total of 356 measurements are performed.

Second, we use a NOSW method to augment the obtained data. Before data augmentation, the total data is split into training set and testing set (with a ratio of $80\%$ : $20\%$). Perform the train/test splitting in advance can ensure that a piece of signal data after any data augmentation would only exit in the training set or the testing set. 

Third, the statistical features are extracted, which include the central trend statistics, the dispersion degree statistics and the distribution shape statistics (see \ref{sec:featureExtracion}). Thus the training feature set and the testing feature set are obtained separately.

Fourth, the adaptive selection feature engineering is applied on both the training feature set and the testing feature set to obtain the optimal feature set. The adaptive selection feature engineering in our methods includes feature aggregation (see \ref{sec:AdaptiveFeatureAggregation}) and feature crosses (see \ref{sec:AdaptiveFeatureCrosses}), both of them contain the adaptive selection modules. The adaptive selection module uses the method of solving the Gini index \cite{gastwirth1972estimation} in the XGBoost algorithm to calculate the importance of the features in the feature data. We can also adaptively select $top-k$ features to form the optimal feature subset according to different data.

Finally, the optimal training feature set and testing feature set are used for the training and prediction of the XGBoost algorithm, respectively. Since the amount of no-flow status is very small and cavitation phenomenon is more relevant for practice, we regard both turbulent-flow and no-flow as ``no cavitation''. Therefore, we study binary classification (``cavitation'' / ``no-cavitation'') and four class cavitation-staging detection.

\section{Results and Discussions}
\label{sec:4-Results and Discussion}
In this section, we evaluate the performance of the proposed cavitation detection framework on the unseen test data. The results of binary classification and four-class classification will be shown. Finally, dependencies on the top $k$ and window-size parameters of our framework will be analyzed.
\subsection{Evaluation Metric}
In order to evaluate the model after training, we selected three metrics to comprehensively assess the model performance: the  Receiver  Operating  Characteristic (ROC) curve, the  Area  Under  Curve (AUC) value, and the Accuracy (Other metrics, see Appendix A). 

\begin{figure}[htbp]
    \centering
    \includegraphics[width=0.45\textwidth,height=45mm]{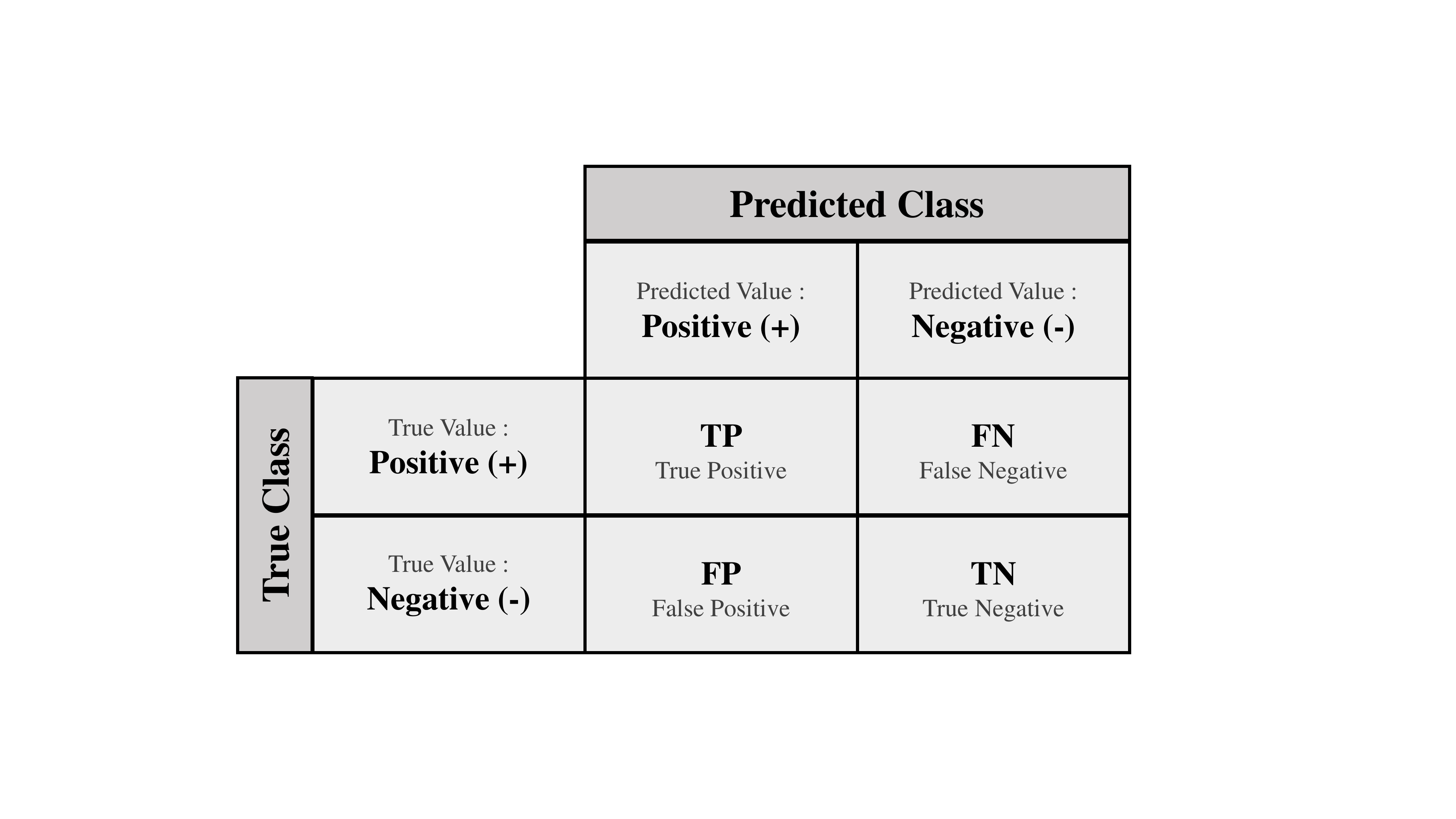}
    \caption{The Confusion Matrix}
    \label{fig:ConfusionMatrix}
\end{figure}

We first calculate the confusion matrix \cite{sokolova2009systematic} to more conveniently define the evaluation metrics and visualize model performance. In the confusion matrix, see Figure \ref{fig:ConfusionMatrix}, each column represents the predicted class, and each row represents the actual class. TP (True Positive) is the fraction of positive samples those got correctly predicted by the model, and TN (True Negative) is for the correctly predicted negative samples. FP (Fasle Positive) means the incorrectly classified positive samples those should be negative actually, and FN (False Negative) is for the incorrectly predicted negative samples.

Accuracy is a common evaluation metric for classification problems, defined as:
\begin{equation}
Accuracy = \frac{{TP + TN}}{{TP + TN + FP + FN}}  
\end{equation}

The ROC \cite{bradley1997use} curve is a line connecting each point with the False Positive Rate (FPR) as the horizontal axis and the True Positive Rate (TPR) as the vertical axis. The definition of FPR and TPR is as follows:
\begin{equation}
FPR = \frac{{FP}}{{FP + TN}} 
\end{equation}
\begin{equation}
TPR = \frac{{TP}}{{FP + FN}}    
\end{equation}

The AUC \cite{bradley1997use} is defined as the area under the ROC curve. In general, the ROC curve is above the straight line $y = x$ and thus $AUC \in [0.5,1]$. Usually the ROC curve and the AUC value are introduced for the binary classification task. In this work, however, by transforming the label we calculate the AUC value and ROC curve in the four-class classification. The specific method is as follows:
\begin{itemize}
    \item Assuming the number of test samples is $j$, the number of classes is $k$.
    \item After training, the probability of each test sample in each class is calculated, and a probability matrix ${P_{jk}}$ is obtained.
    \item The label of each test sample is converted by One-Hot encoding ($0 \to 1000,1 \to 0100,2 \to 0010,3 \to 0001$). Each position is used to mark whether it belongs to the corresponding class. The label matrix ${L_{jk}}$ is obtained.
\end{itemize}

The label matrix ${L_{jk}}$ and probability matrix ${P_{jk}}$ are expanded by rows. Then two columns are obtained after transposition. Finally, the results of binary classification are obtained. Therefore, the final ROC curve and AUC value can be directly obtained by this method. The model results are more intuitively shown through the ROC curve. The larger the AUC value, the better the model performance.

\subsection{Results}
\subsubsection{Results and comparisons with XGBoost}
This subsection focuses on the recognition performance of cavitation detection using our method. In Tabel \ref{tab:NoDataAugmentation} we show the comparison of the cavitation detection performance between the traditional XGBoost method (without NOSW and ASFE) and XGBoost with NOSW plus ASFE. 
\begin{table*}[htbp]
	\caption{Comparison of testing accuracy of our method and traditional XGBoost}
	\label{tab:NoDataAugmentation}
	\centering
	\begin{tabular}{lcccc}
		\toprule
		\multirow{2}{*}{Method}&\multicolumn{2}{c}{\tabincell{c}{Two Classification \\(top $k$ = 9, window size = 2334720)}}&\multicolumn{2}{c}{\tabincell{c}{Four Classification \\(top $k$ = 8, window size = 1556480)}} \\
		\cmidrule(r){2-3}
		\cmidrule(r){4-5}

		& Accuracy   & AUC & Accuracy   & AUC\\
		\midrule
		XGBoost                      & 0.8889  & 0.9218 & 0.8056 & 0.8990             \\
		XGBoost + ASFE + NOSW (our)                   & \textbf{0.9356}  & \textbf{0.9809} & \textbf{0.9167} & \textbf{0.9529}       \\
		\bottomrule
	\end{tabular}
\end{table*}

\begin{figure*}
\centering
\subfigure[Binary Classification]{
\begin{minipage}[t]{0.5\linewidth}
\centering
\includegraphics[width=0.95\textwidth,height=60mm]{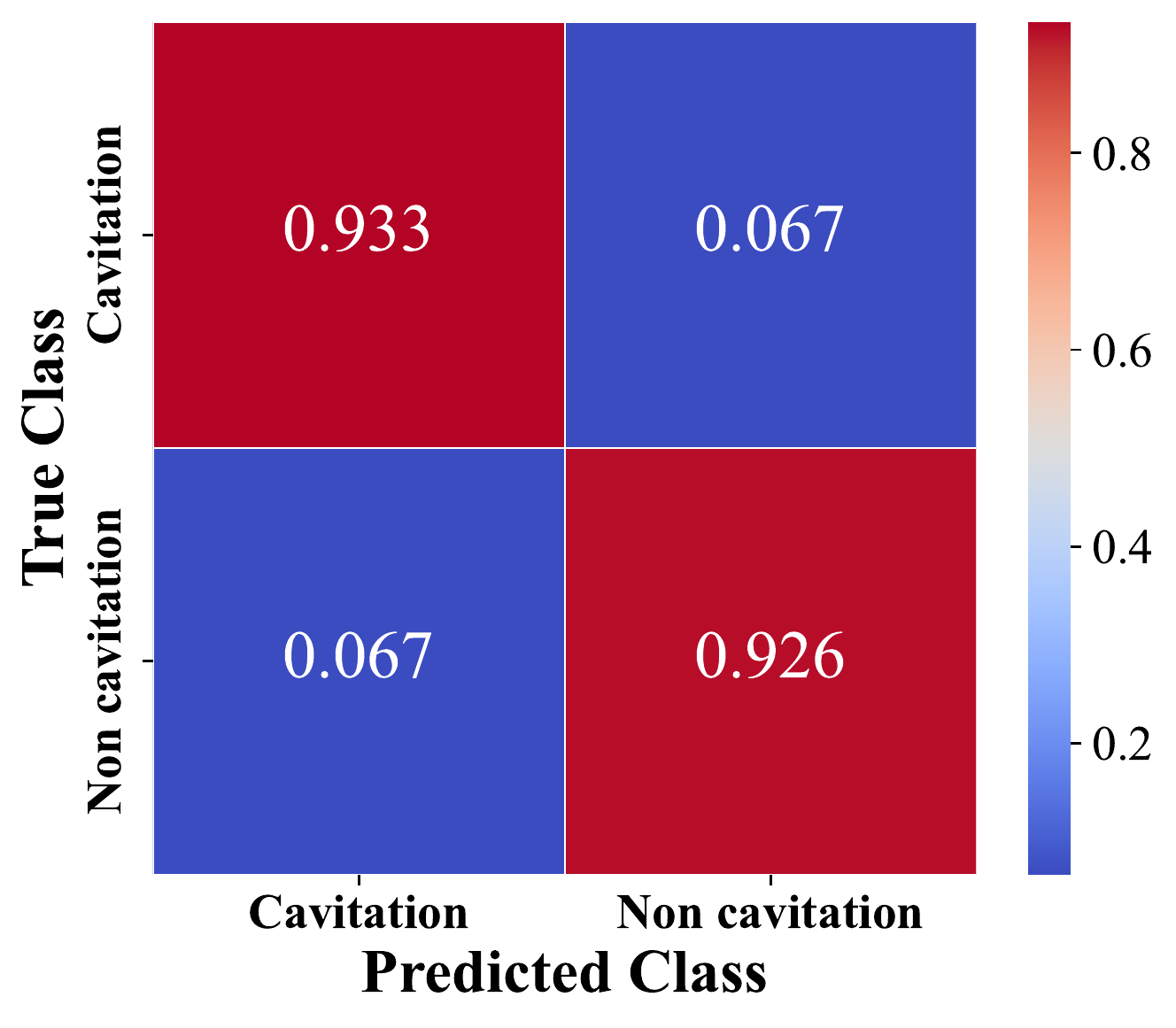}
\end{minipage}%
}%
\subfigure[Four-class Classification]{
\begin{minipage}[t]{0.5\linewidth}
\centering
\includegraphics[width=0.95\textwidth,height=60mm]{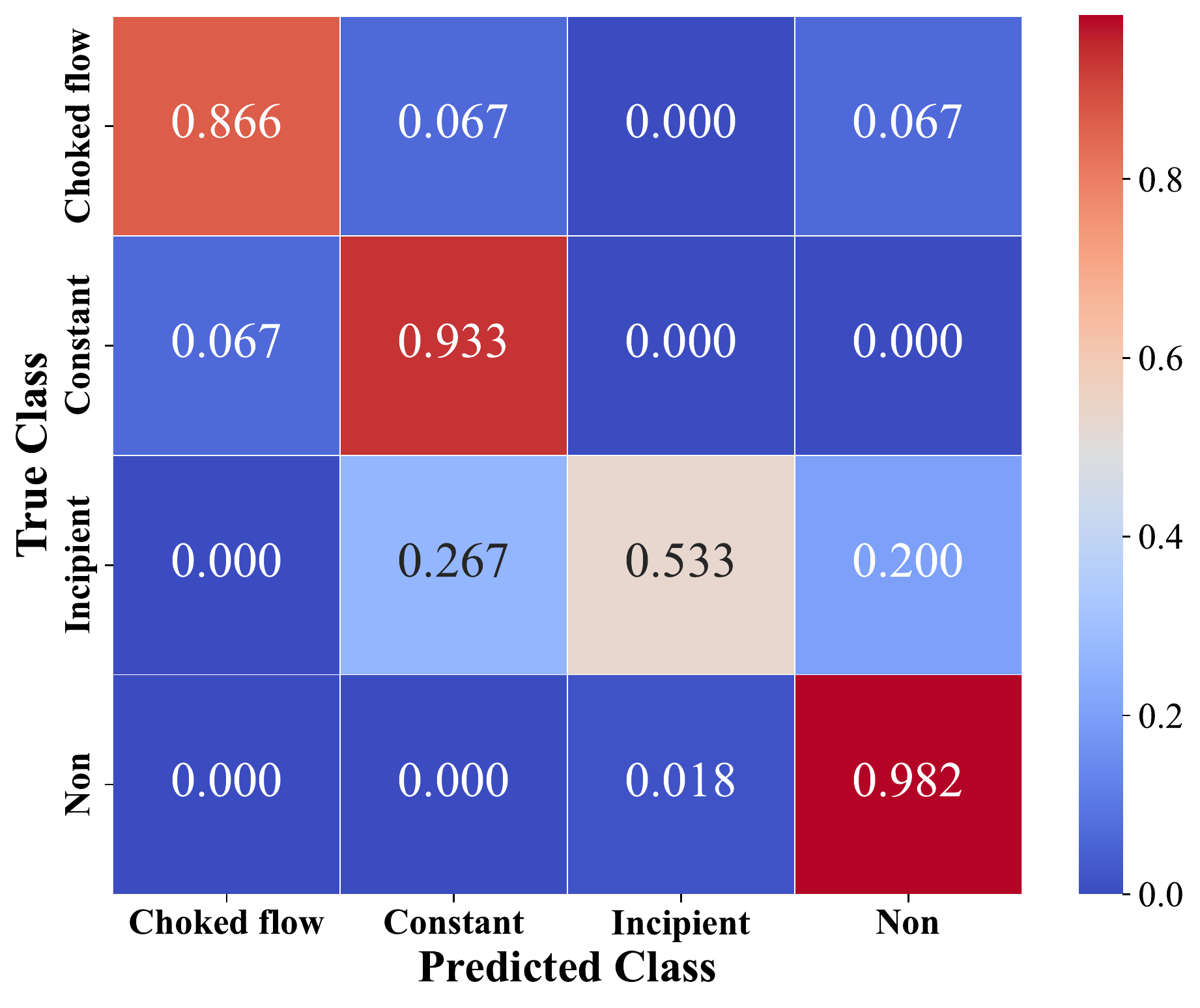}
\end{minipage}%
}%
\centering
\caption{The confusion matrix of our method for binary classification of cavitation/ non-cavitation (a), and four-class classification of choked-flow-cavitation/ contant-cavitation/ incipient-cavitation/ non-cavitation (b).}
\label{fig:Two-Four COnfusion Matrix}
\end{figure*}

For binary classification, the accuracy and AUC value of our method of XGBoost+NOSW+ASFE are increased by 4.67$\%$ and 5.91$\%$, compared to the traditional XGBoost, and for four-class classification the accuracy and AUC value are increased by 11.11$\%$ and 5.39$\%$, compared to the traditional XGBoost method, respectively. These clearly demonstrated the effectiveness of the combination of ASFE and NOSW into XGBoost algorithm to identify the cavitation of flow in valve.

To show the classification results of cavitation detection in detail, we calculate the confusion matrices of our method and show in Figure \ref{fig:Two-Four COnfusion Matrix}. For binary classification, it can be seen that cavitation and no cavitation are easy within our method to be distinguished with a high accuracy. For four-class classification, we see that the constant cavitation is more easily to be recognized with higher accuracy than other cavitation-stages intensity (Cavitation choked flow and incipient cavitation). Comparatively, cavitation choked flow can be recognized with higher accuracy compared to incipient cavitation. Therefore, both choked flow cavitation and constant cavitation are cast as states in the phenomenon of cavitation hysteresis. They have more obvious cavitation characteristics. Being differently, the incipient cavitation is challenging to be identified, because in physics it is the critical state between cavitation and no cavitation. Technically another reason is that the amount of incipient cavitation data is very small. In the study of Lehmann and Young\cite{lehman1964experimental} it also shows that the end stages of cavitation can be more easily detected than incipient cavitation. From a practical perspective, it's acceptable that the incipient cavitation is recognized as the constant cavitation to give rise to process/plant alarm. With this consideration, the accuracy of the incipient cavitation in our method can reach 80$\%$.

\begin{figure*}
\centering
\subfigure[Binary Classification]{
\begin{minipage}[t]{0.5\linewidth}
\centering
\includegraphics[width=0.95\textwidth,height=60mm]{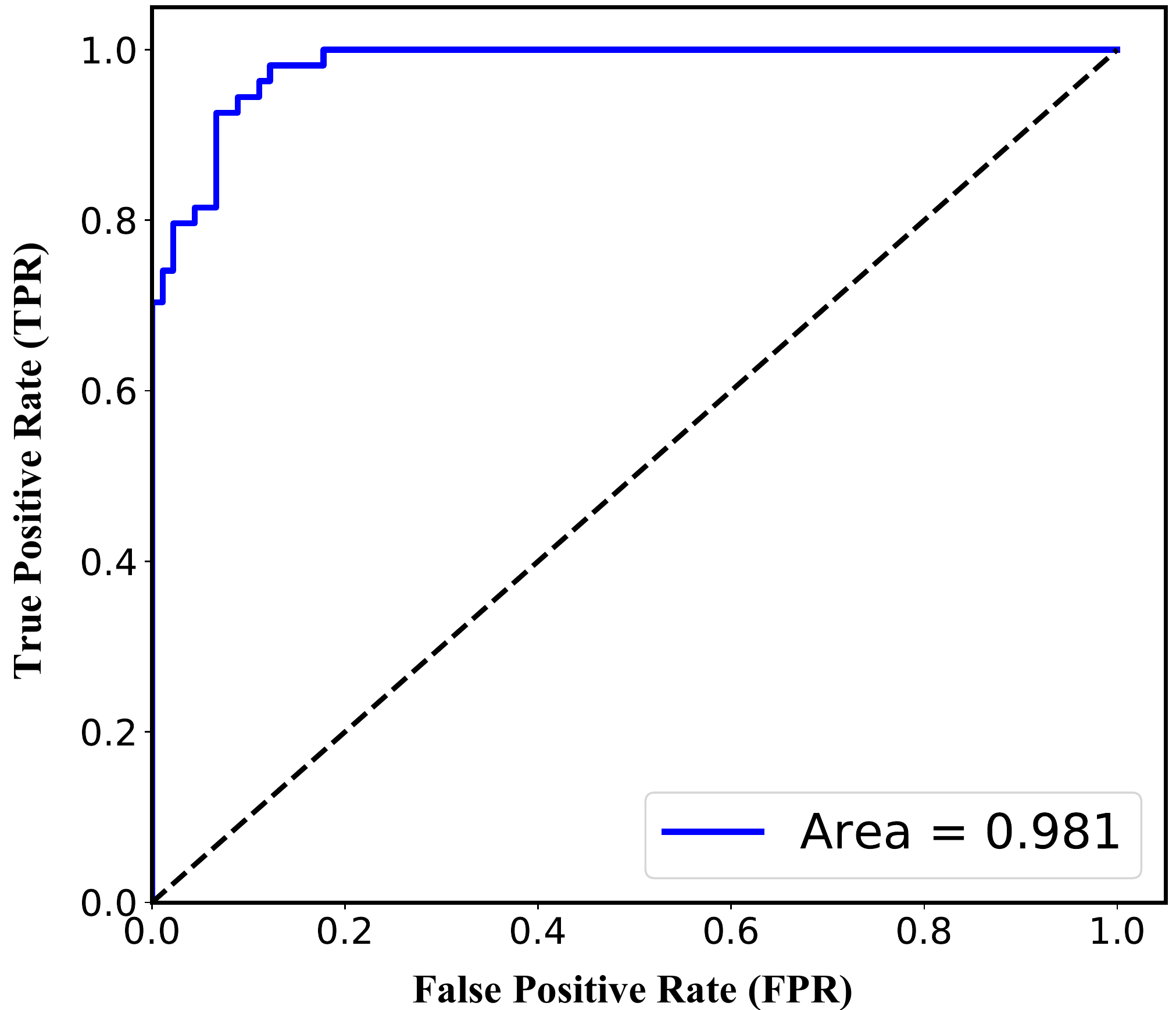}
\end{minipage}%
}%
\subfigure[Four-class Classification]{
\begin{minipage}[t]{0.5\linewidth}
\centering
\includegraphics[width=0.95\textwidth,height=60mm]{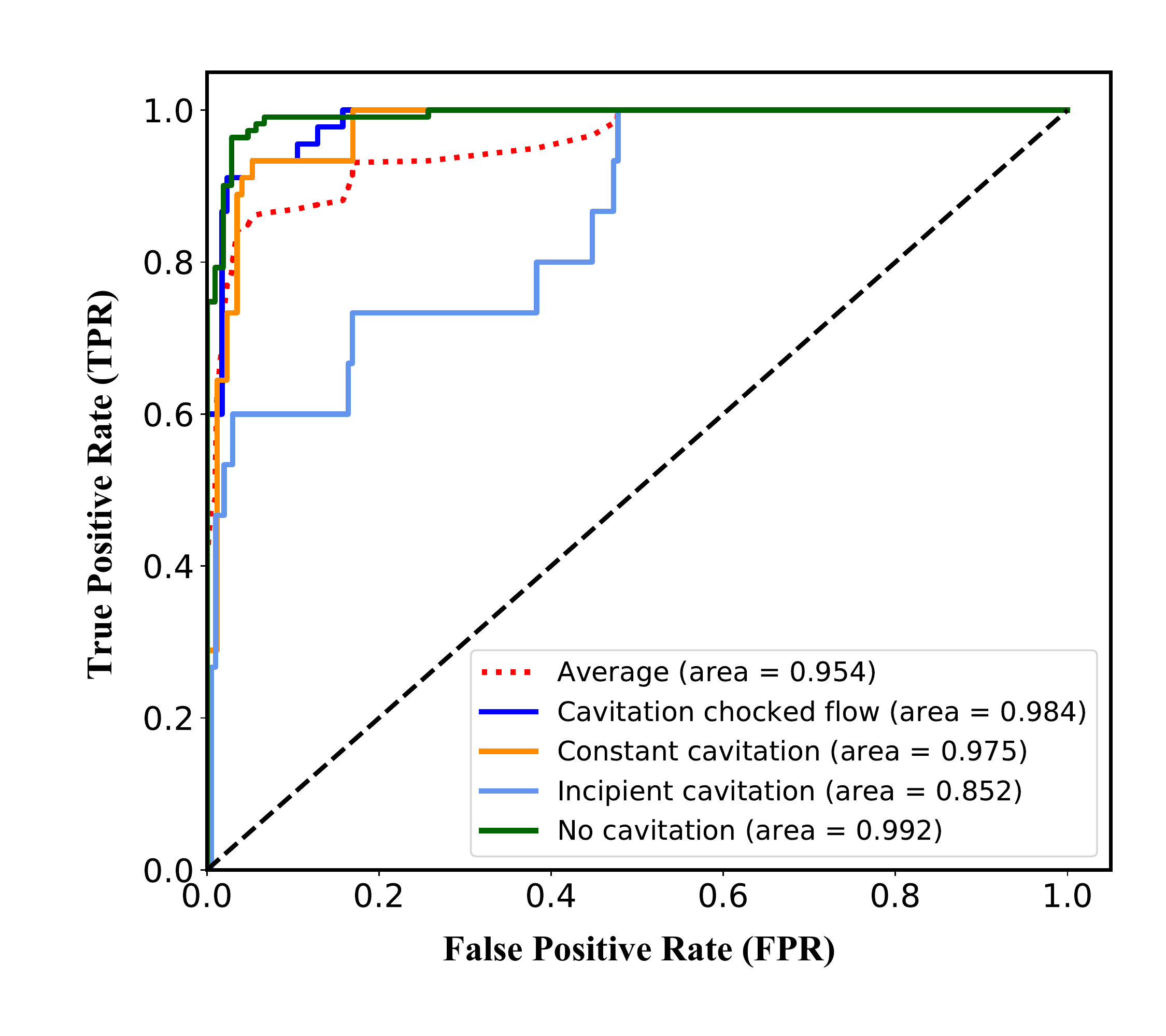}
\end{minipage}%
}%
\centering
\caption{The ROC curve of our method for binary classification of cavitation/ non-cavitation (a), and four-class classification of choked-flow-cavitation/ contant-cavitation/ incipient-cavitation/ non-cavitation (b).}
\label{fig:Two-Four ROC}
\end{figure*}

The Receiver Operator Characteristic (ROC) curve of our method on the testing set is shown in Figure \ref{fig:Two-Four ROC}. For binary classification, the ROC area of our method is 0.981. For four-class classification, the average ROC area of our method for cavitation detection is 0.954. We see that cavitation choked flow, constant cavitation and no-cavitation can be easily identified with higher ROC curve area.
\subsubsection{Parameters analysis}
From a detailed evaluation, the optimal top $k$ value and sliding-window size for the binary classification and four-class classification separately pinned down. The value of top $k$ is chosen to be integers from 5 to 10. The sliding-window size is set to be 2334720, 1556480, 1167360, 933888, 778240, 667062, 583680, 518825 and 466944, respectively. The corresponding results are shown in Table \ref{tab:DifferentKwithDataAugmentation}.
\begin{table}[htbp]
	\caption{Results of Accuracy with different values for $k$ and window size of our method}
	\label{tab:DifferentKwithDataAugmentation}
	\centering
	\small
	\begin{tabular}{ccccccc}
		\toprule
		\multirow{3}{*}{\tabincell{c}{Window \\Size}}&\multicolumn{6}{c}{Two Classification (Accuracy ($\%$))} \\
		\cmidrule(r){2-7}
		& 5 & 6 & 7 & 8 & 9 & 10       \\
		\midrule
		2334720    & 92.36 & 92.36 & 92.36 & 92.36 & \textbf{93.56} & 93.06 \\
		1556480    & 92.59 & 92.13 & 92.13 & 92.59 & 92.59 & 92.13 \\
	    1167360    & 92.01 & 92.36 & 90.97 & 92.01 & 91.32 & 92.36 \\
	    933888     & 91.11 & 91.67 & 91.11 & 90.83 & 90.56 & 90.83 \\
	    778240     & 91.90 & 91.90 & 92.13 & 92.36 & 92.13 & 92.36 \\
	    667062     & 91.07 & 91.27 & 91.27 & 91.07 & 90.87 & 90.67 \\
	    583680     & 90.97 & 91.32 & 91.32 & 91.32 & 91.15 & 91.32 \\
	    518825     & 91.51 & 91.82 & 91.20 & 91.36 & 91.51 & 91.36 \\
	    466944     & 90.56 & 90.83 & 91.25 & 90.83 & 90.56 & 90.69 \\
		\midrule
		\midrule
		\multirow{3}{*}{\tabincell{c}{Window \\Size}}&\multicolumn{6}{c}{Four Classification (Accuracy ($\%$))} \\
		\cmidrule(r){2-7}
		& 5 & 6 & 7 & 8 & 9 & 10       \\
		\midrule
		2334720    & 88.89 & 88.89 & 88.89 & 89.58 & 89.58 & 88.19 \\
		1556480    & 88.89 & 88.89 & 91.20 & \textbf{91.67} & \textbf{91.67} & 91.20 \\
	    1167360    & 80.21 & 83.33 & 80.56 & 80.21 & 80.21 & 80.56 \\
	    933888     & 87.50 & 86.67 & 87.50 & 87.78 & 87.78 & 87.50 \\
	    778240     & 83.10 & 84.72 & 83.10 & 84.26 & 84.95 & 85.19 \\
	    667062     & 85.12 & 85.91 & 87.50 & 86.90 & 85.91 & 86.51 \\
	    583680     & 85.59 & 85.24 & 85.59 & 84.45 & 85.07 & 83.72 \\
	    518825     & 86.42 & 85.19 & 86.88 & 87.19 & 86.27 & 85.80 \\
	    466944     & 86.11 & 87.22 & 86.67 & 85.69 & 85.56 & 85.56 \\
	    \bottomrule
	\end{tabular}
\end{table}
\begin{figure*}
\centering
\subfigure[Binary Classification]{
\begin{minipage}[t]{0.5\textwidth}
\includegraphics[width=0.9\textwidth,height=50mm]{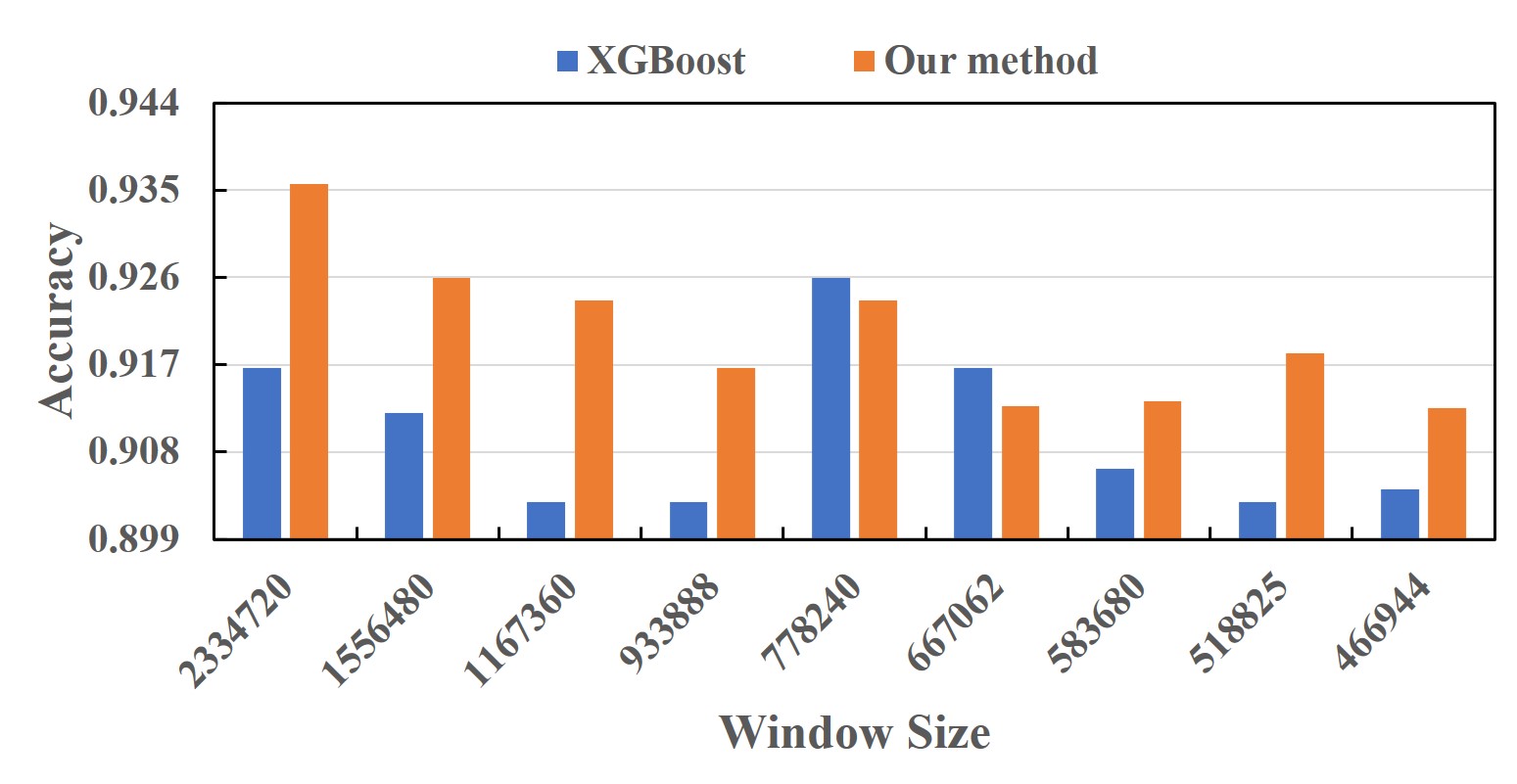}
\end{minipage}%
}%
\subfigure[Four-class Classification]{
\begin{minipage}[t]{0.5\textwidth}
\includegraphics[width=0.9\textwidth,height=50mm]{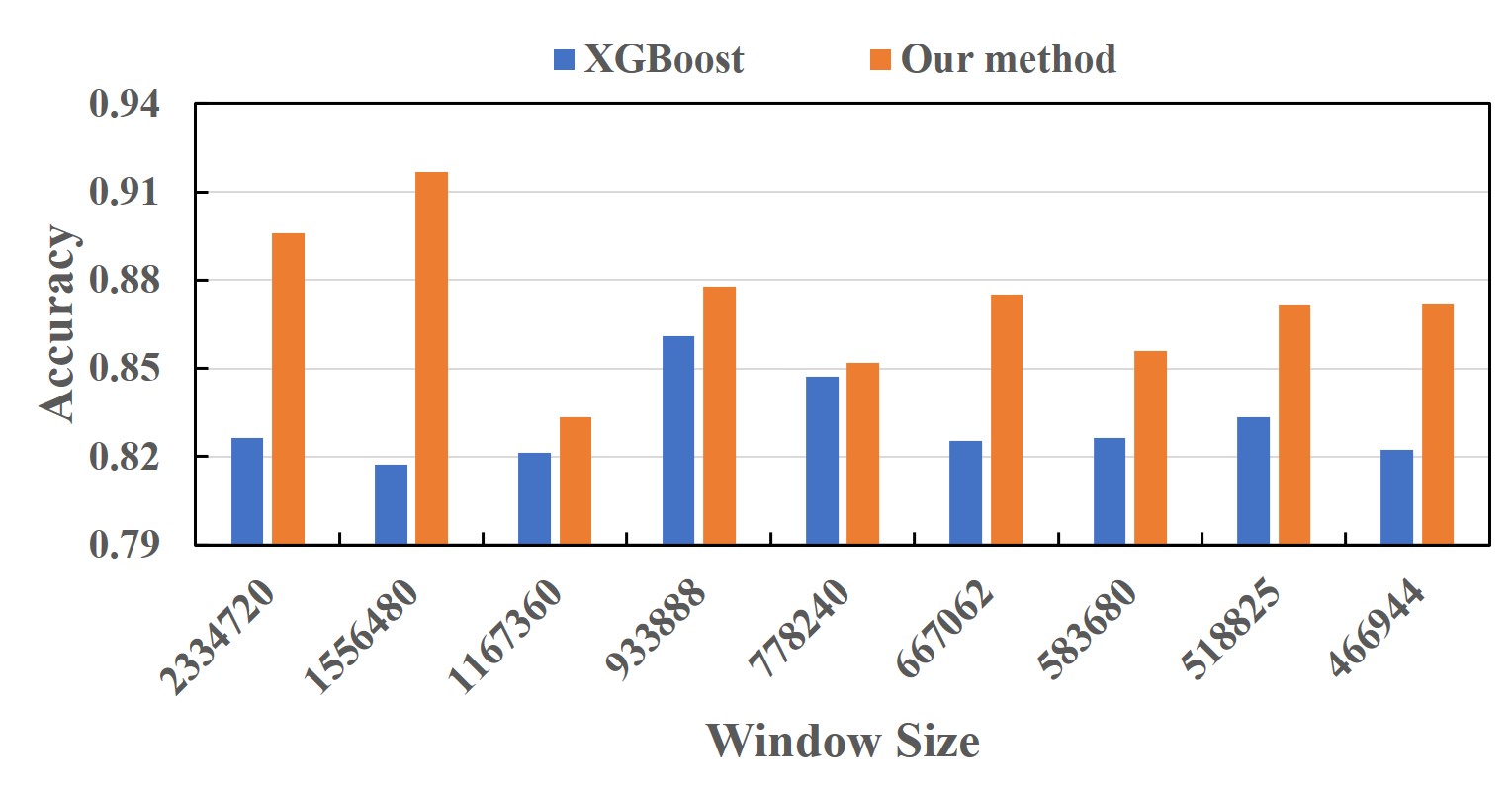}
\end{minipage}%
}%
\caption{Effect of window size on the best accuracy under traditional XGBoost + NOSW and our method.} 
\label{fig:Two-Four With the optiamal k}
\end{figure*}

From Table \ref{tab:DifferentKwithDataAugmentation}, the performance of our method achieves the best for binary classification when $k$ is 9 and window size is 2334720. The value of mean accuracy of our method is \textbf{92.68$\%$}, 92.36$\%$, 91.84$\%$, 91.02$\%$, 92.13$\%$, 91.04$\%$, 91.23$\%$, 91.46$\%$ and 90.79$\%$ with different window size values. Therefore, our method is more stable, when the window size is 2334720. For four-class classification, the performance of our method achieves the best when $k$ is 8 or 9 and the window size is 1556480. The value of mean accuracy of our method is 89.00$\%$, \textbf{90.59$\%$}, 80.85$\%$, 87.46$\%$, 84.22$\%$, 86.31$\%$, 84.94$\%$, 86.29$\%$ and 86.14$\%$ with different window size values. Therefore, our method is more stable, when the window size is 1556480.

And we show the best accuracy of traditional XGBoost + NOSW and our method (XGBoost + NOSW + ASFE) under different window sizes. The results are shown in Figure \ref{fig:Two-Four With the optiamal k}. As shown in Figure \ref{fig:Two-Four With the optiamal k} (a), in almost all window sizes, our method achieves better performances than XGBoost + NOSW. The best result of our method is 93.56$\%$ at the window size is 2334720. The best result of XGBoost + NOSW is 92.59$\%$ at the window size is 778240. The accuracy value  of  our  method  are  increased  by 0.97$\%$, compared with the XGBoost + NOSW. As shown in Figure \ref{fig:Two-Four With the optiamal k} (b), in all window sizes, our method achieves better performances than XGBoost + NOSW. The best result of our method is 91.67$\%$ at the window size is 1556480. The best result of XGBoost + NOSW is 86.11$\%$ at the window size is 778240. The accuracy value of our method are increased  by 5.56$\%$, compared with XGBoost + NOSW. 

\subsubsection{Correlation analysis}
In order to quantitatively analyze correlation between features, we analyze the correlation of different sub-sequences of the same sample. If the correlation of different subsequences is weak, the correlation of statistical features extracted from different sub-sequences is also weak. Conversely, the correction is strong.

\begin{figure*}
\centering
\subfigure[Cavitation Choked Flow]{
\begin{minipage}[t]{0.4\linewidth}
\centering
\includegraphics[width=\textwidth,height=50mm]{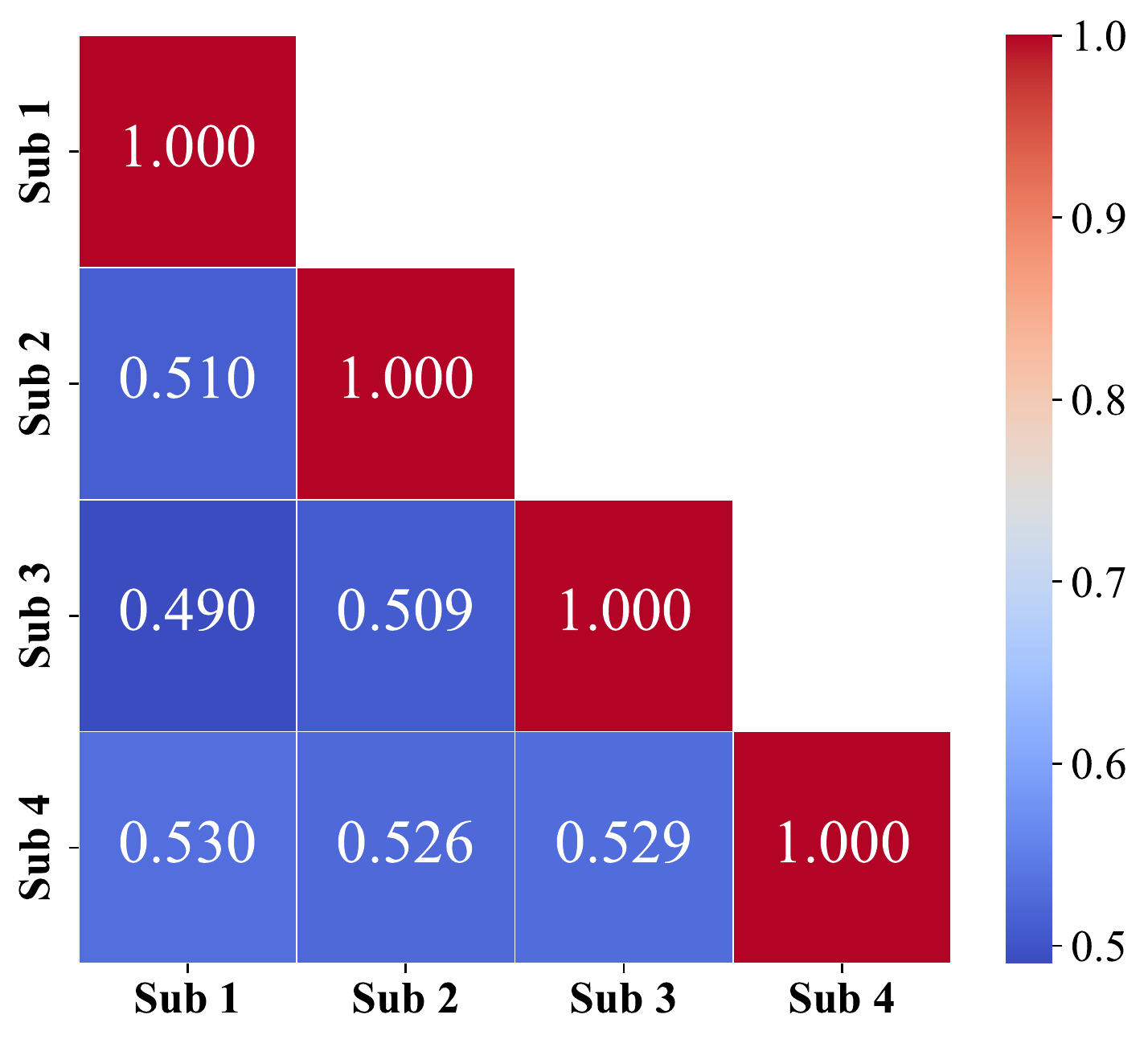}
\end{minipage}%
}%
\subfigure[Constant Cavitation]{
\begin{minipage}[t]{0.4\linewidth}
\centering
\includegraphics[width=\textwidth,height=50mm]{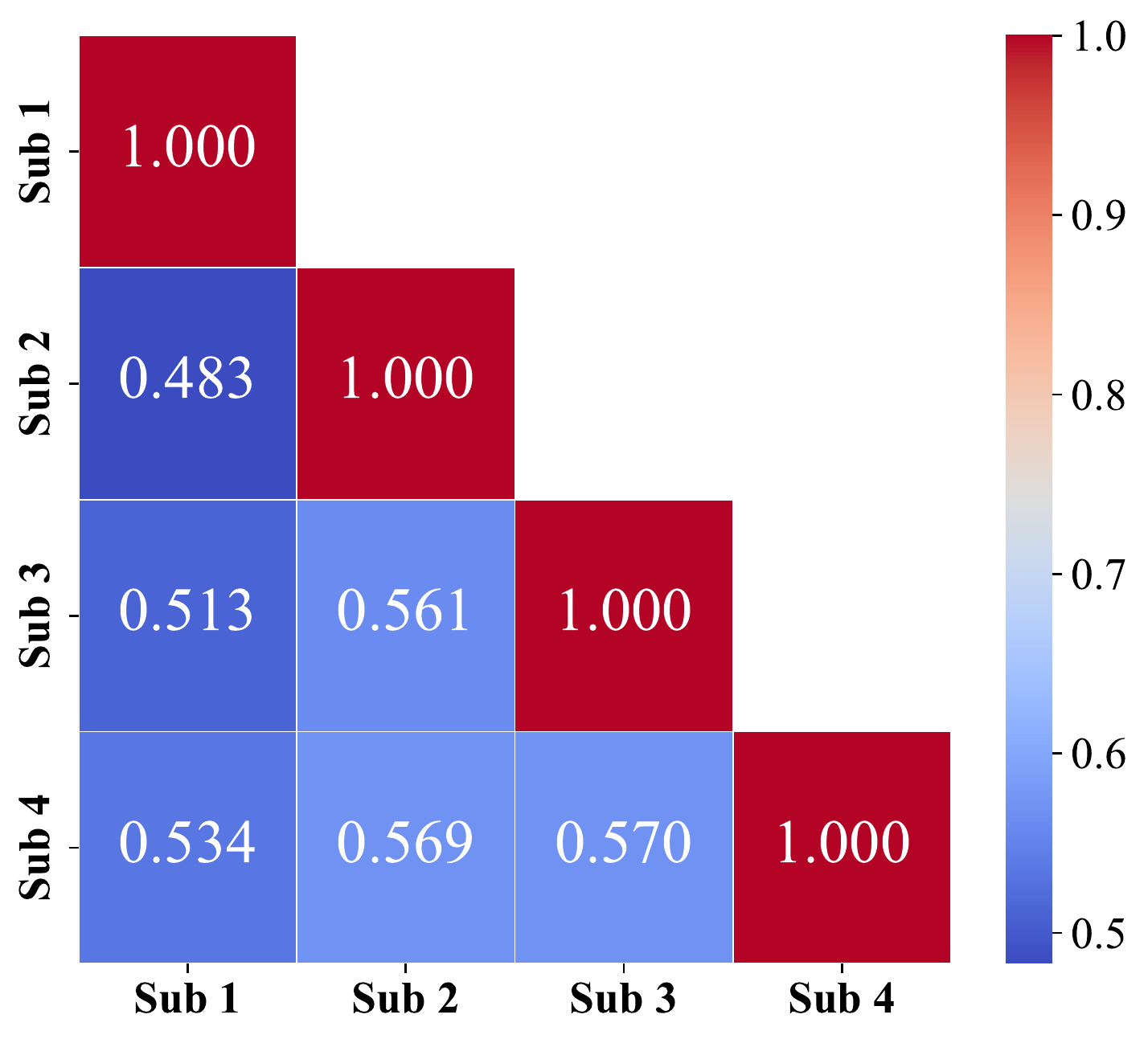}
\end{minipage}%
}%

\subfigure[Incipient Cavitation]{
\begin{minipage}[t]{0.4\linewidth}
\centering
\includegraphics[width=\textwidth,height=50mm]{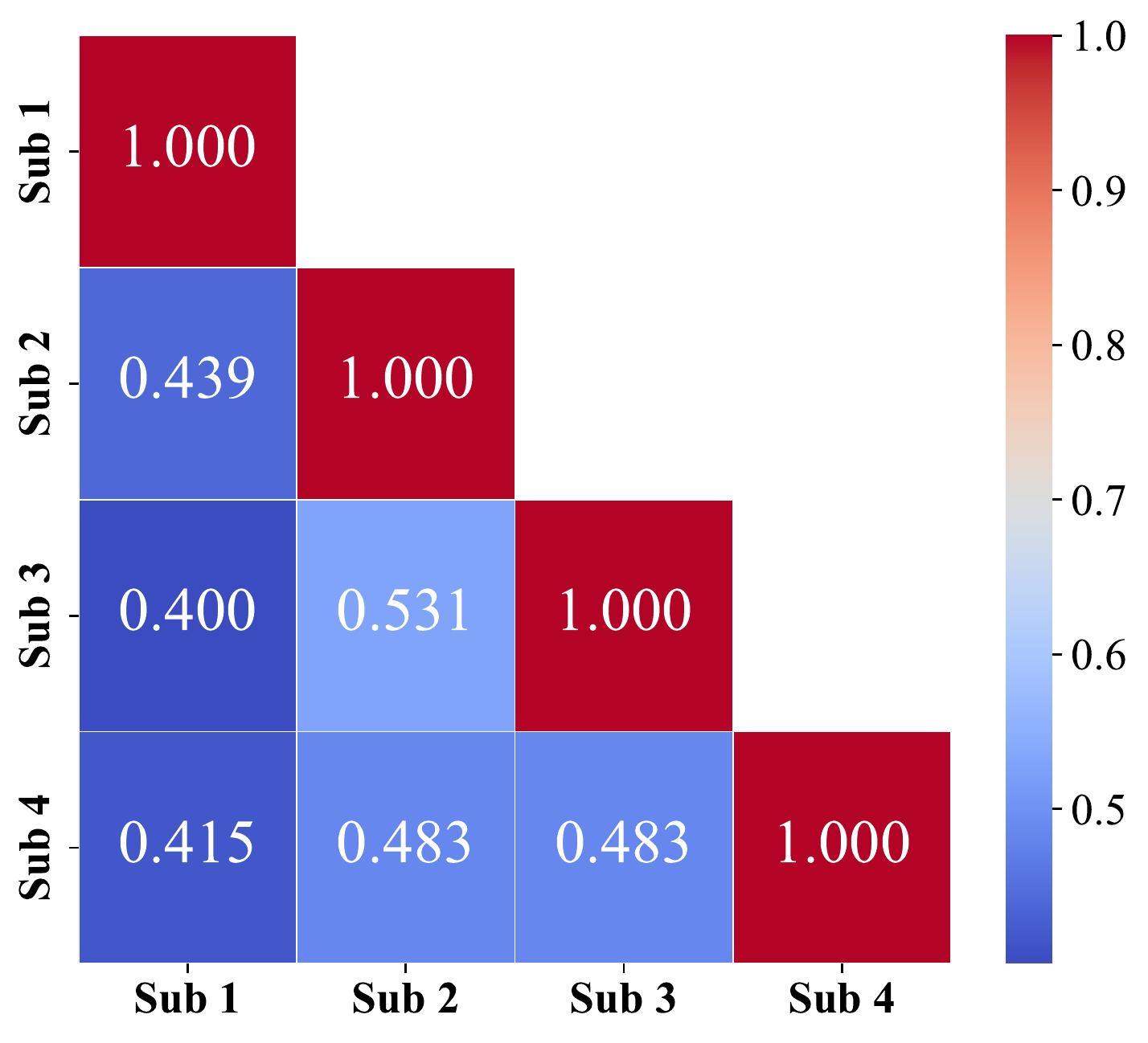}
\end{minipage}%
}%
\subfigure[Non cavitation]{
\begin{minipage}[t]{0.4\linewidth}
\centering
\includegraphics[width=\textwidth,height=50mm]{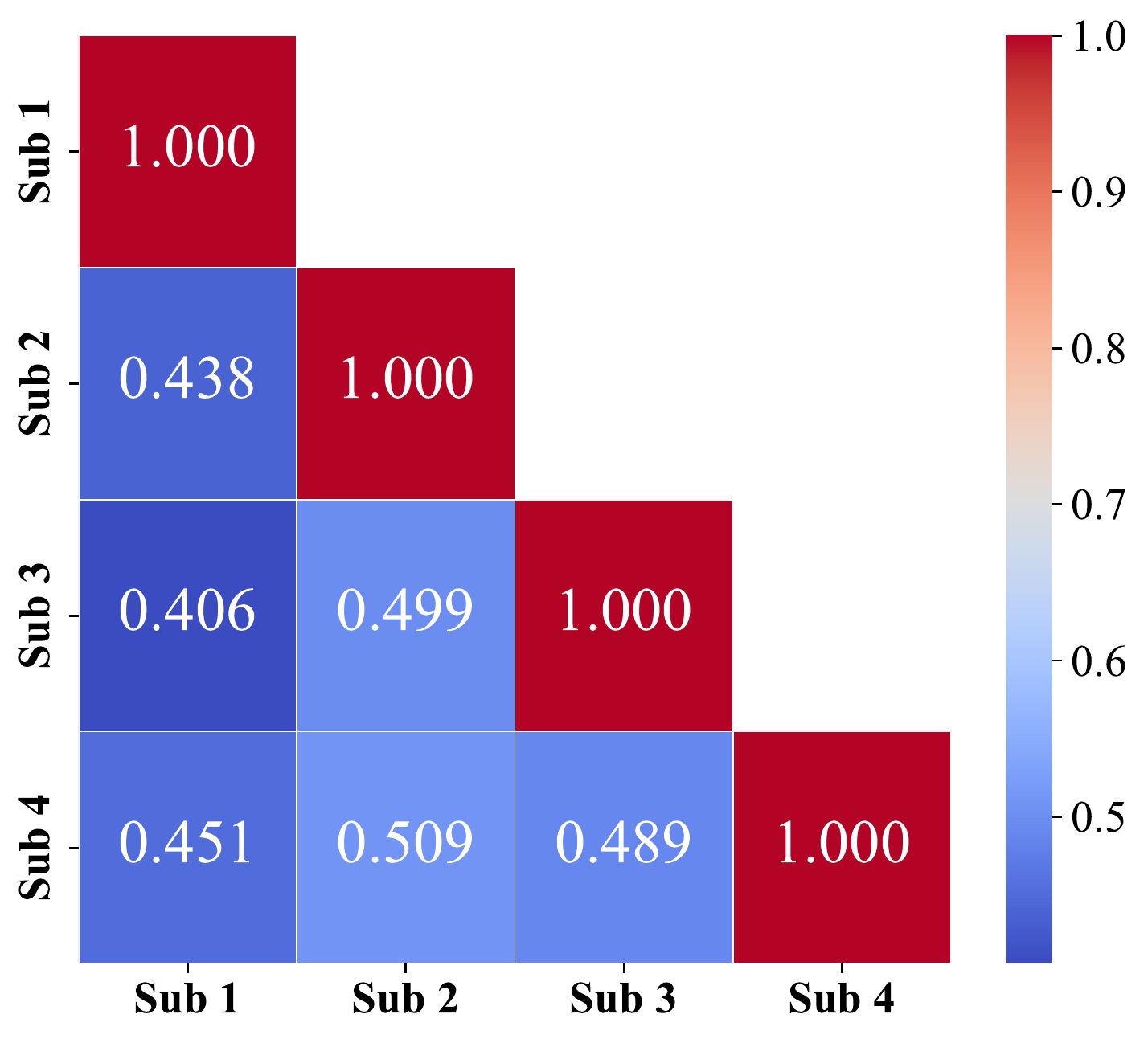}
\end{minipage}%
}%
\centering
\caption{The correlation of different sub-sequences of the same sample for different cavitation states.}
\label{fig:correlation}
\end{figure*}

Figure \ref{fig:correlation} shows the results of Pearson correlation coefficients between different sub-sequences of the same sample for different cavitation states. From Figure \ref{fig:correlation}, it can be seen that the correlations between different sub-sequences are distributed between 0.40 and 0.57. This indicates that the correlation between different sub-sequences presents a moderate degree of correlation for any of the cavitation states. Furthermore, the features extracted from different sub-sequences of the same sample will not be highly correlated, i.e., the features will not be redundant. Since these different sub-sequences represent the same cavitation state, there is no weak or extremely weak correlation between different sub-sequences of the same sample for any cavitation state.

\subsubsection{Comparison results between XGBoost and XGBoost+NOSW}
\begin{figure}[htbp]
    \centering
    \includegraphics[width=0.45\textwidth,height=40mm]{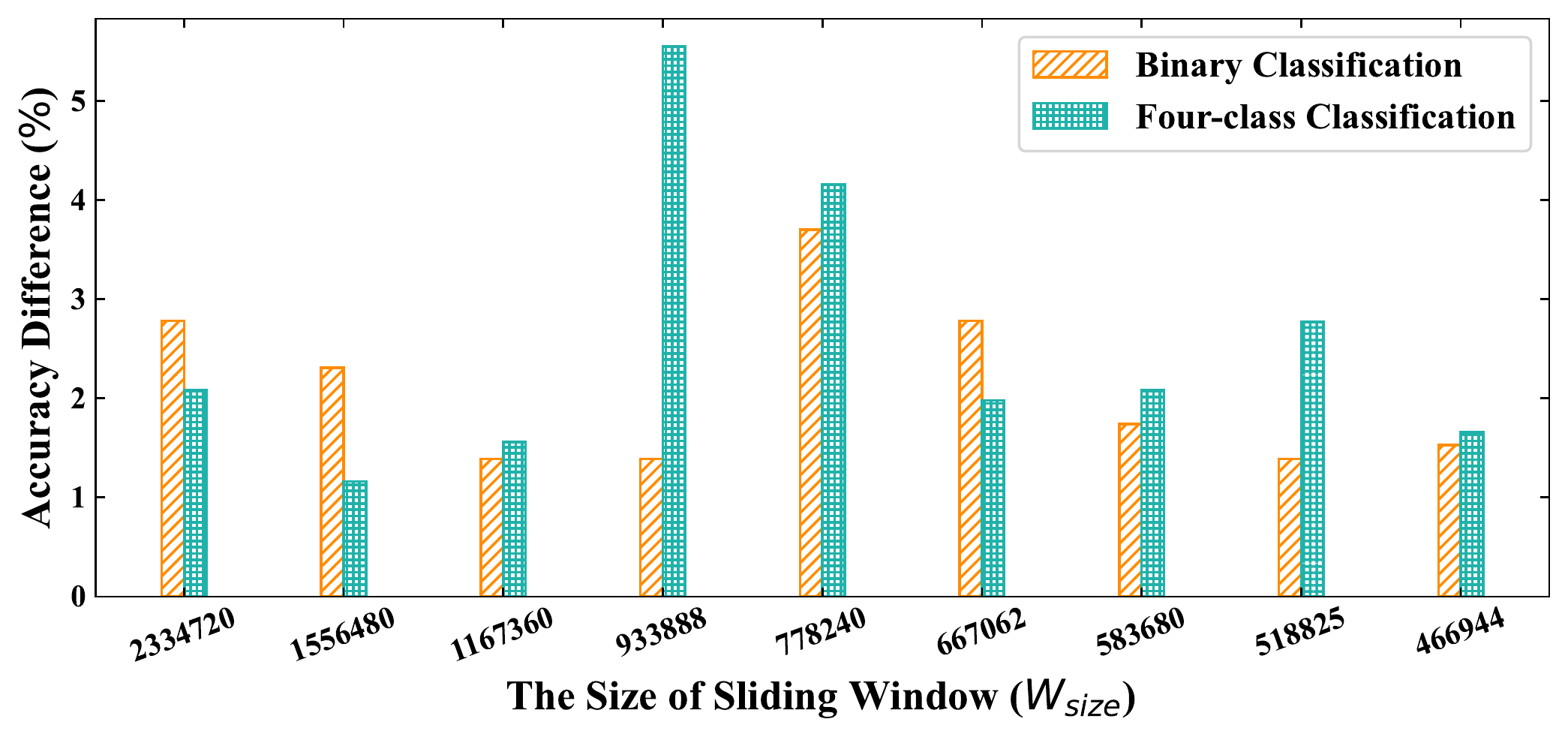}
    \caption{The results of XGBoost and XGBoost+NOSW accuracy difference in binary classification (cavitation detection) and four-class classification (cavitation intensity recognition).}
    \label{fig:results_nosw}
\end{figure}

Figure \ref{fig:results_nosw} shows the accuracy difference between XGBoost on the augmented data and the original data in binary classification (cavitation detection) and four-class classification (cavitation intensity recognition). From figure \ref{fig:results_nosw}, it can be seen that the accuracy difference is above zero at each particular window size, which indicates that the results of XGBoost in the augmented data are better compared to the original data. And the larger the accuracy difference indicates the more significant effect from data augmentation. Furthermore, this implicates that data augmentation using sliding windows is effective. And it also means that the extraction of statistical features after data augmentation does not lead to feature redundancy.

\subsubsection{Comparison results with other methods}
To demonstrate the feasibility of our method, Our approach is compared to traditional machine learning with SVM \cite{yang2005cavitation} and decision trees (DT) \cite{sakthivel2010vibration} and 1D convolutional neural networks (1D CNN) \cite{shervani2018cavitation} for deep learning (see Table \ref{tab:conpared with other methods}). 

For four-class classification (cavitation intensity recognition), XGBoost + NOSW+ASFE gets the best cavitation intensity recognition accuracy of 91.67$\%$ when window size is 1556480. The best cavitation intensity recognition results from SVM, Decision Tree and 1D CNN are 68.71$\%$, 53.86$\%$ and 87.14$\%$, respectively, which are surpassed by 22.96$\%$, 37.81$\%$ and 4.53$\%$ compared to our method. This shows that our method achieves better performance in cavitation intensity recognition than deep learning and traditional machine learning. For binary classification (cavitation detection), XGBoost + NOSW+ASFE achieves the best accuracy 93.56$\%$ when Wsize is 2334720. And the accuracy of the XGBoost + NOSW+ASFE is above 91$\%$ at each value of window size. The best cavitation detection results from SVM and Decision Tree are 83.29$\%$ and 77.00$\%$, respectively, which are surpassed by 10.27$\%$ and 16.56$\%$ compared to our method. Although the results of our method do not outperform the 1D CNN, we only decrease by 0.9$\%$. In addition, the time complexity of our method is much smaller than that of 1D CNN, and our method does not require higher computational hardware.
\begin{table}[htbp]
\centering
\small
\setlength{\tabcolsep}{1mm}{
\caption{The results of different window sizes of binary classification and four-class classification}
\label{tab:conpared with other methods}
\begin{tabular}{cccccc}
\toprule
\multicolumn{2}{c}{\multirow{2}{*}{\begin{tabular}[c]{@{}c@{}}Window Size\\ (W)\end{tabular}}} & \multicolumn{4}{c}{Method}                              \\ 
\cmidrule{3-6} 
\multicolumn{2}{c}{}& SVM   &DT & 1D CNN         & Our method     \\ 
\midrule
\multirow{6}{*}{\begin{tabular}[c]{@{}c@{}}Binary \\ Classification\end{tabular}} 
                                                                & 2334720                      & 81.43 & 73.57         & \textbf{94.29} & 93.56          \\
                                                                & 1556480                      & 81.98 & 75.69         & \textbf{94.36} & 92.59          \\
                                                                & 1167360                      & 82.86 & 76.07         & \textbf{94.29} & 92.36          \\
                                                                & 778240                       & 80.24 & 74.29         & \textbf{94.05} & 92.36          \\
                                                                & 583680                       & 79.82 & 76.96         & \textbf{94.46} & 91.32          \\
                                                                & 466944                       & 83.29 & 77.00         & \textbf{94.14} & 91.25          \\ 
\midrule
\multirow{6}{*}{\begin{tabular}[c]{@{}c@{}}Four-class \\ Classification\end{tabular}}           
                                                                & 2334720                      & 65.71 & 47.86         & 75.71          & \textbf{89.58} \\
                                                                & 1556480                      & 65.42 & 46.58         & 77.48          & \textbf{91.67} \\
                                                                & 1167360                      & 65.00 & 42.86         & 79.64          & \textbf{83.33} \\
                                                                & 778240                       & 64.52 & 50.71         & 80.71          & \textbf{85.19} \\
                                                                & 583680                       & 64.11 & 53.57         & 81.61          & \textbf{85.59} \\
                                                                & 466944                       & 68.71 & 53.86         & 87.14          & \textbf{87.22} \\ 
\bottomrule
\end{tabular}
}
\end{table}

\section{Conclusion and Future Research}
\label{sec:5-Conclusion}
In this paper, the framework of XGBoost combined with adaptive slection feature engineering (ASFE) is proposed to detect cavitation directly using acoustic signal for control valve. Besides to be the first application of XGBoost for cavitation identification, other main contributions of this paper are summarized as the following two points. Firstly, a non-overlapping sliding window data augmentation method is introduced to solve the small-sample learning problem. Secondly, the adaptive selection feature engineering (ASFE) method based on feature scores is proposed to increase the expression ability of original statistical features, Where both the adaptive feature aggregation and feature crosses are used. The proposed ASFE clearly improves the final prediction accuracy for cavitation detection task using XGBoost. Our method is tested on the cavitation data collected within Samson AG and has achieved significant results being superior to the traditional XGBoost method without ASFE. Moreover, the influence of different window size and top $k$ values in our method is detailedly studied. The best combination of window size and top $k$ values suitable for binary classification and four-class classification cavitation detection is pinned down respectively.

The limitations of the proposed method compared to the references may include the following aspects. First, the features for the algorithm are extracted manually. Second, the adaptive selection feature engineering uses physical variables measurements, e.g. the valve opening rate and the upstream pressure. Thirdly, the low recognition of the incipient cavitation. Therefore, in future work, we will explore the end-to-end cavitation detection with solely acoustic signal data using 1D residual network or other convolutional neural networks. In addition, we will also explore Few-shot learning to solve data scarcity and small sample problems.

\section*{Acknowledgements}
\label{sec:acknowledgements}
This research is supported by Xidian-FIAS International Joint Research Center (Y. S.), by the AI grant at FIAS through SAMSON AG (J. F., K. Z.),  by the BMBF funding through the ErUM-Data   project (K. Z.), by SAMSON AG (D. V., T. S., A. W.), by the Walter GreinerGesellschaft zur F\"orderung der physikalischen Grundla-genforschung e.V. through the Judah M. Eisenberg Lau-reatus Chair at Goethe Universit\"at Frankfurt am Main (H. S.), by the NVIDIA GPU grant through NVIDIA Corporation (K. Z.).

\bibliographystyle{unsrt}
\bibliography{references}
\section*{Appendix A}
\label{sec: appendiex}
\setcounter{table}{0}
\setcounter{figure}{0}
\setcounter{equation}{0}
\renewcommand{\thetable}{A\arabic{table}}
\renewcommand{\thefigure}{A\arabic{figure}}
\renewcommand{\theequation}{A.\arabic{equation}}
Precision, Recall and F1-score are common evaluation metric for classification problems, defined as:
\begin{equation}
Precision = \frac{{TP}}{{TP + FP}}
\end{equation}
\begin{equation}
{\mathop{\rm Re}\nolimits} call = \frac{{TP}}{{TP + FN}}
\end{equation}
\begin{equation}
F1-score = \frac{{2 \times \Pr ecision \times {\mathop{\rm Re}\nolimits} call}}{{\Pr ecision + {\mathop{\rm Re}\nolimits} call}}
\end{equation}
where, TP (True Positive) is the positive sample correctly classified by the model. TN (True Negative) is the negative sample correctly classified by the model. FP (Fasle Positive) is the positive sample incorrectly classified by the model. FN (False Negative) is the negative sample incorrectly classified by the model. 

And Precision and Recall are two contradictory and unified metrics. F-score is a weighted summation average of precision and recall.
\subsection*{Binary classification}
The precision, recall and F1-score of XGBoost + NOSW are shown in Table \ref{tab:appendix_xgboost_nosw_two}.
\begin{table}[htbp]
	\caption{Results of Precision, Recall and F1-score with different window size of XGBoost + NOSW}
	\label{tab:appendix_xgboost_nosw_two}
	\centering
	\footnotesize
	\begin{tabular}{cccc}
		\toprule
		\multirow{3}{*}{Window size}&\multicolumn{3}{c}{Evaluation metric ($\%$)} \\
		\cmidrule(r){2-4}
		& Precision & Recall & F1-score        \\
		\midrule
        4669440               & 89.17 & 88.89 & 88.96 \\
		2334720               & 91.92 & 92.59 & 91.72 \\
		1556480               & 91.53 & 91.20 & 91.27  \\
	    1167360               & 90.65 & 90.28 & 90.35  \\
	    933888                & 90.77 & 90.28 & 90.37  \\
	    778240                & 93.24 & 92.59 & 92.67  \\
	    667062                & 92.18 & 91.67 & 91.75  \\
	    583680                & 91.32 & 90.62 & 90.73  \\
	    518825                & 90.77 & 90.27 & 90.37  \\
	    466944                & 90.89 & 90.40 & 90.50  \\
	    \bottomrule
	\end{tabular}
\end{table}

The precision, recall and F1-score of our method (XGBoost + NOSW + ASFE) are shown in Tables \ref{tab:appendix_xgboost_nosw_asfe_precision}, \ref{tab:appendix_xgboost_nosw_asfe_recall} and \ref{tab:appendix_xgboost_nosw_asfe_f1score}.
\begin{table}[htbp]
	\caption{Results of precision with different values for $k$ and window size of our method (XGBoost + NOSW + ASFE)}
	\label{tab:appendix_xgboost_nosw_asfe_precision}
	\centering
	\footnotesize
	\begin{tabular}{ccccccc}
		\toprule
		\multirow{3}{*}{\tabincell{c}{Window \\Size}}&\multicolumn{6}{c}{Precision ($\%$)} \\
		\cmidrule(r){2-7}
		& 5 & 6 & 7 & 8 & 9 & 10       \\
		\midrule
        4669440    & 91.41 & 90.38 & 91.92 & 93.12 & 93.14 & 93.14 \\
		2334720    & 92.52 & 92.52 & 92.52 & 92.52 & 93.14 & 93.14 \\
		1556480    & 92.84 & 92.44 & 92.44 & 92.84 & 92.84 & 92.44 \\
	    1167360    & 92.14 & 92.61 & 91.14 & 92.22 & 91.45 & 92.45 \\
	    933888     & 91.37 & 92.19 & 91.65 & 91.14 & 90.90 & 91.06 \\
	    778240     & 92.12 & 92.01 & 92.27 & 92.48 & 92.27 & 92.47 \\
	    667062     & 91.31 & 91.43 & 91.43 & 91.25 & 90.95 & 90.74 \\
	    583680     & 91.19 & 91.58 & 91.63 & 91.53 & 91.34 & 91.49 \\
	    518825     & 91.88 & 92.19 & 91.49 & 91.62 & 91.84 & 91.71 \\
	    466944     & 90.82 & 91.10 & 91.57 & 91.18 & 91.00 & 91.06 \\
	    \bottomrule
	\end{tabular}
\end{table}
\begin{table}[htbp]
	\caption{Results of recall with different values for $k$ and window size of our method (XGBoost + NOSW + ASFE)}
	\label{tab:appendix_xgboost_nosw_asfe_recall}
	\centering
	\footnotesize
	\begin{tabular}{ccccccc}
		\toprule
		\multirow{3}{*}{\tabincell{c}{Window \\Size}}&\multicolumn{6}{c}{Recall ($\%$)} \\
		\cmidrule(r){2-7}
		& 5 & 6 & 7 & 8 & 9 & 10       \\
		\midrule
        4669440    & 90.28 & 90.28 & 91.67 & 93.06 & 93.56 & 93.06 \\
		2334720    & 92.36 & 92.36 & 92.36 & 92.36 & 93.56 & 93.06 \\
		1556480    & 92.59 & 92.13 & 92.13 & 92.59 & 92.59 & 92.13 \\
	    1167360    & 92.01 & 92.36 & 90.97 & 92.01 & 91.32 & 92.36 \\
	    933888     & 91.11 & 91.67 & 91.11 & 90.83 & 90.56 & 90.83 \\
	    778240     & 91.90 & 91.90 & 92.13 & 92.36 & 92.13 & 92.36 \\
	    667062     & 91.07 & 91.27 & 91.27 & 91.07 & 90.87 & 90.67 \\
	    583680     & 90.97 & 91.32 & 91.32 & 91.32 & 91.15 & 91.32 \\
	    518825     & 91.51 & 91.82 & 91.20 & 91.36 & 91.51 & 91.36 \\
	    466944     & 90.56 & 90.83 & 91.25 & 90.83 & 90.56 & 90.69 \\
	    \bottomrule
	\end{tabular}
\end{table}
\begin{table}[htbp]
	\caption{Results of F1-score with different values for $k$ and window size of our method (XGBoost + NOSW + ASFE)}
	\label{tab:appendix_xgboost_nosw_asfe_f1score}
	\centering
	\footnotesize
	\begin{tabular}{ccccccc}
		\toprule
		\multirow{3}{*}{\tabincell{c}{Window \\Size}}&\multicolumn{6}{c}{F1-score ($\%$)} \\
		\cmidrule(r){2-7}
		& 5 & 6 & 7 & 8 & 9 & 10       \\
		\midrule
		4669440    & 90.40 & 90.31 & 91.72 & 93.08 & 93.08 & 93.08 \\
		2334720    & 92.36 & 92.40 & 92.40 & 92.40 & 93.08 & 93.08 \\
		1556480    & 92.64 & 92.19 & 92.19 & 92.64 & 92.64 & 92.19 \\
	    1167360    & 92.05 & 92.41 & 91.02 & 92.06 & 91.36 & 92.39 \\
	    933888     & 91.17 & 91.75 & 91.20 & 90.90 & 90.63 & 90.89 \\
	    778240     & 91.94 & 91.93 & 92.17 & 92.39 & 92.17 & 92.39 \\
	    667062     & 91.13 & 91.31 & 91.31 & 91.12 & 90.90 & 90.70 \\
	    583680     & 91.02 & 91.38 & 91.38 & 91.37 & 91.20 & 91.36 \\
	    518825     & 91.58 & 91.89 & 91.26 & 91.41 & 91.58 & 91.42 \\
	    466944     & 90.62 & 90.89 & 91.31 & 90.90 & 90.64 & 90.77 \\
	    \bottomrule
	\end{tabular}
\end{table}

\subsection*{Four-class classification}
The precision, recall and F1-score of XGBoost + NOSW are shown in Table \ref{tab:appendix_xgboost_nosw_four}.
\begin{table}[htbp]
	\caption{Results of Precision, Recall and F1-score with different window size of our method (XGBoost + NOSW + ASFE)}
	\label{tab:appendix_xgboost_nosw_four}
	\centering
	\footnotesize
	\begin{tabular}{cccc}
		\toprule
		\multirow{3}{*}{Window size}&\multicolumn{3}{c}{Evaluation metric ($\%$)} \\
		\cmidrule(r){2-4}
		& Precision & Recall & F1-score        \\
		\midrule
        4669440               & 80.01 & 80.56 & 79.68 \\
		2334720               & 83.22 & 82.63 & 81.45 \\
		1556480               & 81.94 & 81.72 & 81.36  \\
	    1167360               & 82.21 & 82.12 & 82.17  \\
	    933888                & 84.90 & 86.11 & 84.81  \\
	    778240                & 85.43 & 84.72 & 82.89  \\
	    667062                & 77.42 & 82.54 & 79.75  \\
	    583680                & 81.25 & 82.64 & 81.67  \\
	    518825                & 80.82 & 83.33 & 81.60  \\
	    466944                & 79.73 & 82.22 & 80.76  \\
	    \bottomrule
	\end{tabular}
\end{table}

The precision, recall and F1-score of our method (XGBoost + NOSW + ASFE) are shown in Tables \ref{tab:appendix_xgboost_nosw_asfe_Precision_four}, \ref{tab:appendix_xgboost_nosw_asfe_Recall_four} and \ref{tab:appendix_xgboost_nosw_asfe_F1score_four}.
\begin{table}[htbp]
	\caption{Results of precision with different values for $k$ and window size of our method (XGBoost + NOSW + ASFE)}
	\label{tab:appendix_xgboost_nosw_asfe_Precision_four}
	\centering
	\footnotesize
	\begin{tabular}{ccccccc}
		\toprule
		\multirow{3}{*}{\tabincell{c}{Window \\Size}}&\multicolumn{6}{c}{Precision ($\%$)} \\
		\cmidrule(r){2-7}
		& 5 & 6 & 7 & 8 & 9 & 10       \\
		\midrule
        4669440    & 87.01 & 88.29 & 86.18 & 87.07 & 86.29 & 84.73 \\
		2334720    & 87.25 & 88.08 & 87.90 & 89.21 & 89.05 & 86.99 \\
		1556480    & 88.51 & 88.56 & 91.04 & 91.47 & 91.47 & 90.86 \\
	    1167360    & 80.03 & 83.28 & 80.24 & 81.24 & 80.43 & 80.86 \\
	    933888     & 86.64 & 85.47 & 87.15 & 87.53 & 87.48 & 87.31 \\
	    778240     & 80.68 & 81.50 & 82.10 & 83.38 & 84.73 & 85.07 \\
	    667062     & 84.88 & 84.44 & 86.55 & 85.73 & 84.91 & 85.32 \\
	    583680     & 84.87 & 84.60 & 85.27 & 83.74 & 84.73 & 84.54 \\
	    518825     & 85.60 & 83.13 & 86.15 & 86.50 & 85.16 & 85.00 \\
	    466944     & 85.40 & 86.99 & 86.02 & 85.15 & 85.17 & 85.21 \\
	    \bottomrule
	\end{tabular}
\end{table}
\begin{table}[htbp]
	\caption{Results of recall with different values for $k$ and window size of our method (XGBoost + NOSW + ASFE)}
	\label{tab:appendix_xgboost_nosw_asfe_Recall_four}
	\centering
	\footnotesize
	\begin{tabular}{ccccccc}
		\toprule
		\multirow{3}{*}{\tabincell{c}{Window \\Size}}&\multicolumn{6}{c}{Recall ($\%$)} \\
		\cmidrule(r){2-7}
		& 5 & 6 & 7 & 8 & 9 & 10       \\
		\midrule
        4669440    & 87.50 & 88.89 & 86.11 & 87.50 & 86.11 & 86.11 \\
		2334720    & 88.89 & 88.89 & 88.89 & 89.58 & 89.58 & 88.19 \\
		1556480    & 88.89 & 88.89 & 91.20 & 91.67 & 91.67 & 91.20 \\
	    1167360    & 80.21 & 83.33 & 80.56 & 80.21 & 80.21 & 80.56 \\
	    933888     & 87.50 & 86.67 & 87.50 & 87.78 & 87.78 & 87.50 \\
	    778240     & 83.10 & 84.72 & 83.10 & 84.26 & 84.95 & 85.19 \\
	    667062     & 85.12 & 85.91 & 87.50 & 86.90 & 85.91 & 86.51 \\
	    583680     & 85.59 & 85.24 & 85.59 & 84.45 & 85.07 & 83.72 \\
	    518825     & 86.42 & 85.19 & 86.88 & 87.19 & 86.27 & 85.80 \\
	    466944     & 86.11 & 87.22 & 86.67 & 85.69 & 85.56 & 85.56 \\
	    \bottomrule
	\end{tabular}
\end{table}
\begin{table}[htbp]
	\caption{Results of F1-score with different values for $k$ and window size of our method (XGBoost + NOSW + ASFE)}
	\label{tab:appendix_xgboost_nosw_asfe_F1score_four}
	\centering
	\footnotesize
	\begin{tabular}{ccccccc}
		\toprule
		\multirow{3}{*}{\tabincell{c}{Window \\Size}}&\multicolumn{6}{c}{F1-score ($\%$)} \\
		\cmidrule(r){2-7}
		& 5 & 6 & 7 & 8 & 9 & 10       \\
		\midrule
		4669440    & 87.05 & 88.14 & 86.02 & 87.23 & 86.18 & 85.12 \\
		2334720    & 87.58 & 88.06 & 88.01 & 89.09 & 89.46 & 87.02 \\
		1556480    & 88.26 & 88.60 & 90.66 & 91.31 & 91.31 & 90.87 \\
	    1167360    & 79.42 & 82.57 & 79.78 & 80.43 & 80.05 & 80.44 \\
	    933888     & 86.55 & 85.89 & 87.22 & 87.53 & 87.52 & 87.27 \\
	    778240     & 81.78 & 82.96 & 82.53 & 83.72 & 84.76 & 85.06 \\
	    667062     & 84.91 & 85.00 & 86.70 & 86.05 & 85.21 & 85.71 \\
	    583680     & 85.14 & 84.83 & 85.39 & 84.08 & 84.87 & 84.59 \\
	    518825     & 85.93 & 83.81 & 86.41 & 86.37 & 85.29 & 85.27 \\
	    466944     & 85.52 & 87.09 & 86.26 & 85.32 & 85.33 & 85.35 \\
	    \bottomrule
	\end{tabular}
\end{table}
\setcounter{table}{0}
\setcounter{figure}{0}
\setcounter{equation}{0}
\renewcommand{\thetable}{A\arabic{table}}
\renewcommand{\thefigure}{A\arabic{figure}}
\renewcommand{\theequation}{A.\arabic{equation}}

\end{document}